\documentclass[10pt,preprint]{aastex}

\newcommand{\vv}[1]{{\bf #1}}

\newcommand{\avg}[1]{{\langle{#1}\rangle}}
\newcommand{\Avg}[1]{{\left\langle{#1}\right\rangle}}
\def\simless{\mathbin{\lower 3pt\hbox
	{$\,\rlap{\raise 5pt\hbox{$\char'074$}}\mathchar"7218\,$}}} 
\def\simgreat{\mathbin{\lower 3pt\hbox
	{$\,\rlap{\raise 5pt\hbox{$\char'076$}}\mathchar"7218\,$}}} 

\newcommand{\band}[2]{\ensuremath{^{{#1}}\!{#2}}}

\newcommand{\kversion}{{\tt v1\_11}}

\newcounter{thefigs}
\newcommand{\fignum}{\arabic{thefigs}}

\newcounter{thetabs}

\newcounter{address}

\slugcomment{To be submitted to \aj}

\shortauthors{Blanton {\it et al.} (2000)}
\shorttitle{Modeling Galaxy SEDs}

\begin{document}
 

\title{Estimating Fixed Frame Galaxy Magnitudes in the SDSS$^1$}



\author{
Michael R. Blanton\altaffilmark{\ref{NYU}},
J.~Brinkmann\altaffilmark{\ref{APO}},
Istv\'an Csabai\altaffilmark{\ref{JHU}},
Mamoru Doi\altaffilmark{\ref{Tokyo}},
Daniel Eisenstein\altaffilmark{\ref{Arizona}},
Masataka Fukugita\altaffilmark{\ref{CosmicRay},\ref{IAS}},
James E. Gunn\altaffilmark{\ref{Princeton}},
David W. Hogg\altaffilmark{\ref{NYU}}, and
David J. Schlegel\altaffilmark{\ref{Princeton}}
}

\altaffiltext{1}{Based on observations obtained with the
Sloan Digital Sky Survey} 
\setcounter{address}{2}
\altaffiltext{\theaddress}{
\stepcounter{address}
New York University, Department of Physics, 4 Washington Place, New
York, NY 10003
\label{NYU}}
\altaffiltext{\theaddress}{
\stepcounter{address}
Department of Physics and Astronomy, The Johns Hopkins University,
Baltimore, MD 21218
\label{JHU}}
\altaffiltext{\theaddress}{
\stepcounter{address}
Department of Astronomy and Research Center for 
the Early Universe,
School of Science, University of Tokyo,
Tokyo 113-0033, Japan
\label{Tokyo}}
\altaffiltext{\theaddress}{
\stepcounter{address}
Steward Observatory, 
933 N. Cherry Ave., Tucson, AZ
85721
\label{Arizona}}
\altaffiltext{\theaddress}{
\stepcounter{address}
Princeton University Observatory, Princeton,
NJ 08544
\label{Princeton}}
\altaffiltext{\theaddress}{
\stepcounter{address}
Apache Point Observatory,
2001 Apache Point Road,
P.O. Box 59, Sunspot, NM 88349-0059
\label{APO}}
\altaffiltext{\theaddress}{
\stepcounter{address}
Institute for Cosmic Ray Research, University of
Tokyo, Midori, Tanashi, Tokyo 188-8502, Japan
\label{CosmicRay}}
\altaffiltext{\theaddress}{
\stepcounter{address}
Institute for Advanced Study, Olden Lane,
Princeton, NJ 08540
\label{IAS}}

\clearpage

\begin{abstract}
Broad-band measurements of flux for galaxies at different redshifts
measure different regions of the rest-frame galaxy spectrum. Certain
astronomical questions, such as the evolution of the luminosity
function of galaxies, require transforming these inherently
redshift-dependent magnitudes into redshift-independent quantities. To
prepare to address these astronomical questions, investigated in
detail in subsequent papers, we fit spectral energy distributions
(SEDs) to broad band photometric observations, in the context of the
optical observations of the Sloan Digital Sky Survey (SDSS). Linear
combinations of four spectral templates can reproduce the five SDSS
magnitudes of all galaxies to the precision of the
photometry. Expressed in the appropriate coordinate system, the
locus of the coefficients multiplying the templates is planar,
and in fact nearly linear. The resulting reconstructed SEDs can be
used to recover fixed frame magnitudes over a range of redshifts. This
process yields consistent results, in the sense that within each
sample the intrinsic colors of similar type galaxies are nearly
constant with redshift. We compare our results to simpler
interpolation methods and galaxy spectrophotometry from the SDSS.  The
software that generates these results is publicly available and easily
adapted to handle a wide range of galaxy observations.
\end{abstract}


%
%

\section{Motivation}
\label{motivation}

In the future, observations of galaxies (and indeed of any
astronomical sources) will be performed using devices which combine
high spatial resolution and high spectral resolution. By that time,
the phenomena of broad band filters and the quaint terminology
surrounding their usage --- magnitudes, $K$-corrections, color terms,
{\it etc.} --- will have long since been forgotten. However, until
such time as wide field spatially-resolved spectroscopy is cheaply
available, virtually all observations of galaxies will be made through
broad band filters, and special care has to be taken to recover
knowledge of galaxy spectral energy distributions (SEDs) from these
observations.  Reconstructing galaxy SEDs is nontrivial because SEDs
of galaxies contain important information on scales considerably
smaller than the width of typical broad band filters. Furthermore,
SEDs of distant galaxies are redshifted such that the more distant the
galaxy, the further the observed bandpass is blueshifted relative to
the rest-frame spectral energy distribution of the object observed.
This paper focuses on a solution to these problems which (in the
optical wavelength regime) is fast, robust, and consistent over a
large range of redshifts.

In the past, people have accounted for these effects using
``$K$-corrections'' applied to the observed magnitudes. In these
analyses, a function $K(z)$ is added to the standard cosmological
bolometric distance modulus $\mathrm{DM}(z)$ to obtain the
relationship between the apparent magnitude $m_R$ of band $R$ and the
absolute magnitude $M$ of (in general different) band $Q$:
\begin{equation}
\label{kcorrecteqn}
m_R = M_Q + \mathrm{DM}(z) + K_{QR}(z)
\end{equation}
The traditional definition of the $K$-correction takes $Q=R$; however,
we note that in practice many surveys do perform $K$-corrections from
one bandpass to another when comparing high redshift and low redshift
observations. 

Sometimes a single function $K(z)$ has been applied regardless of the
galaxy type, though more recently it has become standard to use a
discrete set of $K(z)$ functions depending on galaxy type (based
either on morphology or spectral features). As described in
\cite{oke68a} and later papers, the function is based on the
projection of an assumed galaxy SED redshifted to $z$ onto the
measured instrument response as a function of wavelength (including
the effects of the atmosphere, the reflectivity of the mirrors, the
filter transmission, and the response of the CCD device or
photographic emulsion). For example, for the Sloan Digital Sky Survey
(SDSS) filter system, \citet{frei94a} and \citet{fukugita95a} have
both presented results for the $K$-corrections of galaxies. Each of
these papers tabulates $K(z)$ (called $k(z)$ by \citealt{frei94a}) for
galaxies of various morphological types, based on galaxy
spectrophotometry.\footnote{Note that the description on p. 57 of
\citet{binney98a} of the meaning of $k(z)$ in \citet{frei94a} is
incorrect; in fact, their $k(z)$ is exactly equivalent to our $K(z)$.}

However, galaxies do not all have the same SED, nor are they selected
from some discrete set of SEDs. For this reason, these schemes for
applying the $K$-corrections can fail to be self-consistent: the SED
assumed for the $K$-corrections can be significantly inconsistent with
the observed galaxy colors! As we demand more precision from our
astronomical data analysis in new, large multi-band surveys, and in
particular as we try to quantify the evolution of galaxies, we must
take a more sophisticated approach to approximating fixed frame
observations of galaxies.

Our approach here is to use a method for inferring the underlying SEDs
of a set of galaxies at a range of redshifts by requiring that their
SEDs all be drawn from a similar population. For each galaxy, we will
recover a model SED, which can be used to synthesize the galaxy's
magnitude in any bandpass. Although the operation we are performing on
the magnitudes is not a ``$K$-correction'' in the historical sense of
the term, we will refer to it as such in this and subsequent
papers. Our approach is equivalent (nearly identical) to the
photometric redshift estimation methods of \citet{csabai00a},
except we use slightly different coordinate systems for our SED
template space. In fact, the software includes a fast and relatively
accurate photometric redshift estimator based on their method.

We implemented this system and are publishing it in order to lay the
groundwork for upcoming papers which will rely heavily on the
reliability of the fixed frame magnitudes determined here.  Because
the SDSS is one of the largest astronomical surveys to date, our
method will be useful to a large number of other investigators, both
inside and outside the SDSS collaboration.  Thus, this paper also
serves the purpose of describing the release of a piece of
software. The computer software we distribute builds itself into a C
shared object library, around which we have written both stand-alone C
programs and IDL routines. We have made the source code available
publicly, through a web page and through a public CVS
repository. Improvements or ports to other languages implemented by
users may be incorporated into the code upon request. The conditions
of use for the code are that this paper be cited in any resulting
refereed journal article and that the version of the code used is
specified in any such paper. The version of the code used to make the
figures for this paper is {\tt kcorrect \kversion}.

We refer throughout this paper to $AB$ magnitudes, first defined by
\citet{oke83a} to measure the ratio of the number of photons included
in the signal of the detector relative to that number for a flat
spectrum source with $g(\nu)=3.631 \times 10^{-20}$ ergs cm$^{-2}$
s$^{-1}$ Hz$^{-1}$. For a source with a spectrum $f(\nu)$ the $AB$
magnitude should be (for a perfectly calibrated $AB$ system)
\begin{eqnarray}
\label{ABdef}
m_{AB} &=& -2.41 - 2.5 \log_{10}\left[
\frac{\int_{0}^{\infty} d\lambda \lambda f(\lambda) R(\lambda)}
{\int_{0}^{\infty} d\lambda \lambda^{-1} R(\lambda)}\right]
\cr
&=& -48.60 - 2.5 \log_{10}\left[
\frac{\int_{0}^{\infty} d\nu \nu^{-1} f(\nu) R(\nu)}
{\int_{0}^{\infty} d\nu \nu^{-1} R(\nu)}\right],
\end{eqnarray}
where $R(\lambda)$ is the fraction of photons entering the Earth's
atmosphere which are included in the signal as a function of
wavelength (a unitless quantity). Note that $R(\lambda)$ can be
defined even for devices which do not count photons directly (such as
bolometers). This equation is written such that $f(\lambda)$ is in
units of ergs cm$^{-2}$ s$^{-1}$ \AA$^{-1}$ and $f(\nu)$ is in units
of ergs cm$^{-2}$ s$^{-1}$ Hz$^{-1}$, while $\lambda$ is expressed in
\AA\ and $\nu$ is expressed in Hz. The normalizations defined here
mean that an object with $f(\nu) = g(\nu) = 3631\mathrm{~Jy} = 3.631
\times 10^{-20}$ ergs cm$^{-2}$ s$^{-1}$ Hz$^{-1}$ has all its AB
magnitudes equal to zero.  The $\lambda^{-1}$ appears in the integrand
of the denominator of the first equation because
$g(\lambda)=c/\lambda^2$ for a ``flat spectrum'' source with
$g(\nu)=1$.  The difference in the zeropoints of the two equations
simply corresponds to the factor of the speed of light $c$ (expressed
in \AA\ s$^{-1}$) in that expression for $g(\lambda)$.

Section \ref{sedfit} describes our method of fitting galaxy SEDs to
broad band photometry and how to calculate $K$-corrections. Section
\ref{data} applies the method to galaxies in the SDSS, showing that
the fits are robust.  Section \ref{conclusions} concludes and
discusses future development of the method described here.

\section{Calculating Fixed Frame Galaxy Magnitudes}
\label{sedfit}

First, we describe how we reconstruct galaxy SEDs from broad band
magnitude measurements. Second, we describe how we convert these SEDs
into estimates of fixed frame galaxy magnitudes. 

\subsection{Fitting SEDs to Galaxy Broad-band Magnitudes}

Our task is to recover a model for the galaxy SED from broad band
photometric measurements. Since a set of broad band galaxy fluxes does
not correspond uniquely to a particular SED, and because galaxy SEDs
are known to have significant structure over wavelength ranges small
compared to our bandpass, the inverse problem of reconstructing SEDs
from broad band fluxes is ill-posed. The task is made yet more
difficult by the fact that the separation of the filters are large
compared to their widths; the SEDs are thus not well-sampled in the
Nyquist sense of that term. However, we are not completely ignorant
about the forms which galaxy SEDs take, so we can attempt to use what
we know about galaxy SEDs to appropriately regularize our fits.  The
method described here for doing so follows closely the method of
estimating photometric redshifts used by \citet{csabai00a}.  It is
designed to take advantage of what we already understand about galaxy
spectral energy distributions.

Begin with an SED space defined by $N_b$ template galaxy SEDs
$\vv{v}_i(\lambda)$ (for example from \citealt{bruzual93a}), where $N_b$ is
large. In principle, this should be a complete set of galaxy SEDs if
we want {\it all} galaxy SEDs to be in the space spanned by the
$\vv{v}_i(\lambda)$; in practice, this property is not necessary.
Given a set of $\vv{v}_i(\lambda)$, one can find an orthonormal set of
spectra $\vv{b}_i(\lambda)$ spanning the same space.  To be more
precise, we define the dot-product in the space of galaxy spectra to
be:
\begin{equation}
\vv{v}_i \cdot \vv{v}_j =
\int_{\lambda_{\mathrm{min}}}^{\lambda_{\mathrm{max}}} d\lambda
v_i(\lambda) v_j(\lambda),
\end{equation}
where $\lambda_{\mathrm{min}}$ and $\lambda_{\mathrm{max}}$ define the
wavelength range used to define orthogonality. These limits are
defined not over the full range covered by the $\vv{v}_i(\lambda)$
spectra, but instead over a smaller range which is within observable
wavelengths over most of the redshift range of the sample.\footnote{
If the template spectra were identical in the subrange chosen and only
differed outside of it, our choice would cause computational problems,
but in fact the spectra are all independent in the restricted, as well
as the full, wavelength range.}  We then define the basis of the SED
space such that the $N_b$ template SEDs, $\vv{v}_i$, are linear
combinations of $N_b$ basis spectra $\vv{b}_i$ and that $\vv{b}_i
\cdot \vv{b}_j = \delta_{ij}$ (where $\delta_{ij}$ represents the
Kronecker delta). These conditions do not fully specify the
$\vv{b}_i$, which naturally may be rotated arbitrarily within our
$N_b$-dimensional subspace of spectral space.

In our case the number of observed bands is less than $N_b$, so we
cannot determine an individual galaxy's position in this
$N_b$-dimensional space from its broad band colors alone.  However, as
outlined by \citet{csabai00a}, one can find a lower dimensional space
defined by $N_t$ vectors $\vv{e}_k$ which best fits the {\it set} of
galaxies, as follows. The SED of galaxy $i$ may be reconstructed from a
linear combination of the basis spectra in this lower dimensional
space:
\begin{equation}
\label{lincomb}
f_{\mathrm{rec},i}(\lambda) d\lambda = 
\sum_k a_{i,k} \sum_j e_{k,j} b_j(\lambda) d\lambda,
\end{equation}
where $a_k$ are the components of the galaxy projected in the
low-dimensional space. From this SED it is easy to determine the
reconstructed flux
$F_{\mathrm{rec},il}$ in each survey bandpass by projecting
the model SED onto the bandpass $l$: 
\begin{equation}
F_{\mathrm{rec},il} = \int_0^\infty d\lambda
f_{\mathrm{rec},i}(\lambda) S_l(\lambda),
\end{equation}
where $S_l(\lambda)$ is the response of the instrument in band $l$.  We define
\begin{equation}
\chi^2 = \sum_i \sum_l
\frac{(F_{\mathrm{obs},il}-F_{\mathrm{rec},il})^2}{\sigma_{il}^2},
\end{equation}
where $\sigma_{il}$ is the estimated error in the observed flux
$F_{\mathrm{obs},il}$ in bandpass $l$ for galaxy $i$. Taking
derivatives of $\chi^2$ with respect to $a_{i,k}$ and $e_{k,j}$
results in a set of equations bilinear in $a_{i,k}$ and $e_{k,j}$.
\citet{csabai00a} describe how to iteratively solve this bilinear
equation by first fixing $e_{k,j}$ and fitting for $a_{i,k}$, then
fixing $a_{i,k}$ and fitting for $e_{k,j}$, and iterating that
procedure.


We define the ``total flux'' $f_k$ of each template as the flux in the
range $\lambda_{\mathrm{min}}<\lambda<\lambda_{\mathrm{max}}$; then an
estimate of the flux in this range is $F_{{t}}=\Sigma_k a_k f_k$.
Galaxies of some fixed $F_t$ clearly lie on a plane in the component
space $a_k$. Therefore, a natural choice for the orientation of the
axes $\vv{e}_k$ is to let $\vv{e}_0$ be perpendicular to the planes of
constant $F_t$; in this manner, the coefficient $a_0$ in Equation
(\ref{lincomb}) is directly proportional to $F_t$ and is thus purely a
measure of the object's flux (in the wavelength range
$\lambda_{\mathrm{min}} <\lambda < \lambda_{\mathrm{max}}$), while the
parameters $a_i/a_0$ (where $i>0$) primarily measure the shape of the
galaxy SED in the wavelength range $\lambda_{\mathrm{min}} <\lambda <
\lambda_{\mathrm{max}}$. $F_t$ can be trivially related to $L_t$, the
total luminosity in this range, by the inverse square of the
luminosity distance (from, for example, \citealt{hogg99a}).  As it
turns out, the parameters $a_i/a_0$ tend to be distributed
approximately in an ellipsoid, with not much curvature within the
space defined by the template SEDs. Thus, we transform the axes such
that $a_i/a_0 = 0$ is the mean of the distribution and rotate the axes
$\vv{e}_i$ (for $i>0$) such that they are aligned with the principal
axes of the ellipsoid. Furthermore, we normalize the vectors
$\vv{e}_i$ such that the zeroth component $a_0=F_t$ is output by the
code in units of ergs cm$^{-2}$ s$^{-1}$ (within the range
$\lambda_{\mathrm{min}} <\lambda < \lambda_{\mathrm{max}}$).

In this way, we can reconstruct SEDs based on the broad band
magnitudes of galaxies. From these reconstructed SEDs one can estimate
$K$-corrections, develop a measure of galaxy spectral type, or
synthesize other measurements of galaxy flux.  The templates
determined during the procedure can, of course, be used to estimate
photometric redshifts, as \citet{csabai00a} describes.

\subsection{Applying Direct Constraints on $a_i/a_0$}
\label{direct}

Sometimes we will want to apply direct constraints on the coefficients
$a_i/a_0$ of the galaxy we are fitting. For example, in this paper we
fit our templates to a set of galaxies with high quality photometry in
all bands. However, we would like to use these templates to fit for
the SEDs of other galaxies, which might have much worse photometry in
one or more bands. Since the errors in the photometry can be
catastrophic (that is, non-gaussian), when we fit the SEDs of the
lower signal-to-noise galaxies it makes sense to at least weakly
constrain these galaxies to have reasonable SEDs, in the sense that
the $a_i/a_0$ ought not to be too different than those for
well-measured galaxies. We can assume a gaussian prior probability
distribution with a mean $\avg{a_{c,j}/a_{c,0}}$, a variance matrix
$V_{c,jl}$, and a weight $A_c$.  We calculate a preliminary value of
${\hat{a}}_{c,0}$, assuming $a_i = 0$. Then we constrain our fit by
adding a term to $\chi^2$ as follows:
\begin{equation}
\sum_j \sum_l A_c V^{-1}_{jl}
\left(\frac{a_l}{{\hat{a}}_{c,0}}-\Avg{\frac{a_{c,l}}{a_{c,0}}}\right)
\left(\frac{a_j}{{\hat{a}}_{c,0}}-\Avg{\frac{a_{c,j}}{a_{c,0}}}\right)
\end{equation}
The resulting $\chi^2$ can still be minimized linearly, so we have
sacrificed very little in speed.  Note that with such a constraint,
one can calculate $K$-corrections even for galaxies with only one or
two bands measured, and still obtain sensible results. 

\subsection{Calculating $K$-corrections}

To characterize $K$-corrections, consider again Equation
(\ref{kcorrecteqn}): 
\begin{equation}
m_R = M_Q + \mathrm{DM}(z) + K_{QR}(z).
\end{equation}
The definition of an apparent AB magnitude is given by Equation
(\ref{ABdef}). An absolute AB magnitude is defined to be the apparent
AB magnitude which would be measured for a galaxy 10 pc distant, at
rest.  In the restricted case of AB magnitudes, the resulting
$K$-correction is given by:
\begin{equation}
K_{QR}(z) = -2.5 \log_{10} \left[
\frac{
\int_0^{\infty} \frac{d\nu}{\nu} f(\nu) R(\nu)
}{
\int_0^{\infty} \frac{d\nu}{\nu} f(\nu/(1+z))/(1+z) Q(\nu)
}
\right]
\end{equation}
A more general expression holds for other magnitude systems (such as
the Vega-relative magnitude system). 

In practice, we will be concerned here primarily with transforming the
apparent magnitudes in band $R$ to a band $Q$ which is just the $R$
band shifted by a factor $(1+z_0)$, which we will denote
\begin{equation}
\band{z_0}{R}(\nu) \equiv R(\nu/(1+z_0))  
\end{equation}
read, ``$R$-band shifted to $z_0$.''  

In the case of transforming $R$ to $\band{z_0}{R}$
\begin{equation}
K_{\band{z_0}{R}R}(z) = -2.5 \log_{10} \left[
\frac{
\int_0^{\infty} \frac{d\nu}{\nu} f(\nu) R(\nu)
}{
\int_0^{\infty} \frac{d\nu}{\nu} f(\nu/(1+z))/(1+z) R(\nu/(1+z_0))
}
\right]
\end{equation}
In the special case that $z=z_0$, one is transforming the magnitude of
an observed galaxy at $z_0$ to its rest-frame magnitude in a band
shifted by $z_0$. For that case
\begin{equation}
K_{\band{z_0}{R}R}(z_0) = -2.5 \log_{10} (1+z_0)
\end{equation}
independent of $f(\nu)$ (which is natural, since both magnitudes
sample the same region of $f(\nu)$). The fact that one's uncertainties
in $f(\nu)$ do not matter makes the use of the fixed frame magnitudes
shifted to the redshift of the object in question useful, when it is
possible. We note in passing that \citet{oke68a} state that the
$K$-correction ``would disappear if intensity measurements of
redshifted galaxies were made with a detector whose spectral
acceptance band was shifted by $1+z$ at all wavelengths;'' here we
have just shown this statement to be untrue for an AB system (though
one could define a magnitude system in which this were true).

In practice, one can calculate the fixed frame magnitudes in one of
two ways:
\begin{enumerate}
\item Simply reconstruct $\band{z_0}{R}$ from the model
$f_{\mathrm{rec}}(\nu)$ 
\item Calculate the $K$-correction above and apply it to the observed
$R$ magnitude
\end{enumerate}
If the reconstructions of the observed magnitudes $R$ are very similar
to the actual observations (as we will show is the case for our fits),
there is very little difference between these two methods. However, if
the reconstructions are poor, there will be large differences between
the methods; in particular, method (1) will yield a color distribution
of galaxies with artificially low dimensionality. Having estimated
$\band{z_0}{m}$, you can then calculate the luminosity by simply
applying to the calculated magnitude the cosmological distance modulus
(that is, the luminosity distance-squared law, as tabulated by, {\it
e.g.}, \citealt{hogg99a}).

We want to emphasize here that, while $K$-correcting to a fixed frame
bandpass is {\it sometimes} necessary in order to achieve a scientific
objective, it is not {\it always} necessary or appropriate. Because
$K$-corrections are inherently uncertain (the broad band magnitudes do
not uniquely determine the SED) they should be avoided or minimized
when possible. For example, if one were calculating the evolution of
clustering of red and blue galaxies separately, it would perhaps be
wise to perform the separation not on the $K$-corrected colors of the
galaxies but on the median color of $M_\ast$ galaxies as a function of
redshift. Similarly, in situations where $K$-corrections cannot be
ignored, such as the calculation of the evolution of the luminosity
function, their effect should be minimized by, for example, correcting
to a band shifted to the median redshift of galaxies in the sample.

Typically one's measurements will be difficult to connect to solar
luminosities in $\band{z}{b}$ (unless $z=0$), simply because nobody
has projected the appropriate stellar spectrophotometry onto these
bandpasses. However, it is our position that stars are better
understood than galaxies, and that it is therefore simpler in the end
to stay as close as possible to the system in which the galaxies are
observed. In any case, astronomy is quickly reaching a level of
precision for which the exact nature of the bandpasses used has to be
known and considered in most analyses of observational data.

The reader may ask why, if we are basing our $K$-corrections on a full
model of the galaxy SED, we do not correct to bandpasses with simpler
shapes (for example, top hats). The reason is that we want the
effective bandpasses to actually have been observed for some galaxies
in the sample, so that the $K$-correction for those galaxies are
independent of the galaxy SED.  This property is desirable because, as
noted above, there are inherent uncertainties in the $K$-corrections
due to our lack of knowledge of the SEDs. Nevertheless, we note that
synthesizing top hat magnitudes may be appropriate in certain
situations, as they have the virtue (similarly to the $AB$ magnitude
system) that they are very easy to comprehend and synthesize from
theory.

\subsection{Public Access to the Code}

The version of the $K$-correction code (\kversion) implementing the
calculations described here, along with eigentemplates and filter
curves, is publicly available on the World Wide Web at {\tt
http://physics.nyu.edu/\~\ mb144/kcorrect}.  The whole of the code can
be used through the Research Systems, Incorporated, IDL language;
everything except for the template-fitting also exists in stand-alone
C programs (which call the same routines, guaranteeing consistency).

\section{Application to SDSS Data}
\label{data}

In this section we describe how we applied the above method to the
SDSS data set.

\subsection{SDSS Spectroscopic Data}

The SDSS (\citealt{york00a}) is producing imaging and spectroscopic
surveys over $\pi$ steradians in the Northern Galactic Cap. A
dedicated 2.5m telescope (Siegmund {\it et al.}, in preparation) at
Apache Point Observatory, Sunspot, New Mexico, images the sky in five
bands between 3000 and 10000 \AA\ ($u$, $g$, $r$, $i$, $z$;
\citealt{fukugita96a}) using a drift-scanning, mosaic CCD camera
(\citealt{gunn98a}), detecting objects to a flux limit of $r'\sim
22.5$. The ultimate goal is to spectroscopically observe Main Sample
galaxies 900,000 galaxies (down to $r_{\mathrm{lim}}\approx 17.77$;
\citealt{strauss02a}), 100,000 Luminous Red Galaxies (LRGs;
\citealt{eisenstein01a}), and 100,000 QSOs (\citealt{fan99a}; Newberg
{\it et al.}, in preparation).  This spectroscopic follow up uses two
digital spectrographs on the same telescope as the imaging
camera. Many of the details of the galaxy survey are described in a
description of the galaxy target selection paper
(\citealt{strauss02a}). Other aspects of the survey are described in
the Early Data Release (EDR; \citealt{stoughton02a}).

The results of the photometric pipeline for all of the galaxies were
extracted from the SDSS Operational Database. The photometry used here
for the bulk of these objects was the same as that used when the
objects were targeted. However, for those objects which were in the
EDR photometric catalog, we used the better calibrations and
photometry from the EDR. The versions of the SDSS pipeline PHOTO used
for the reductions of these data ranged from {\tt v5\_0} to {\tt
v5\_2}. The treatment of relatively small galaxies, which account for
most of our sample, did not change substantially throughout these
versions. As described in \citet{smith02a}, magnitudes are calibrated to a
standard star network approximately in the $AB$ system, described in
Section \ref{motivation}.

\begin{figure}
\figurenum{\fignum}
\plotone{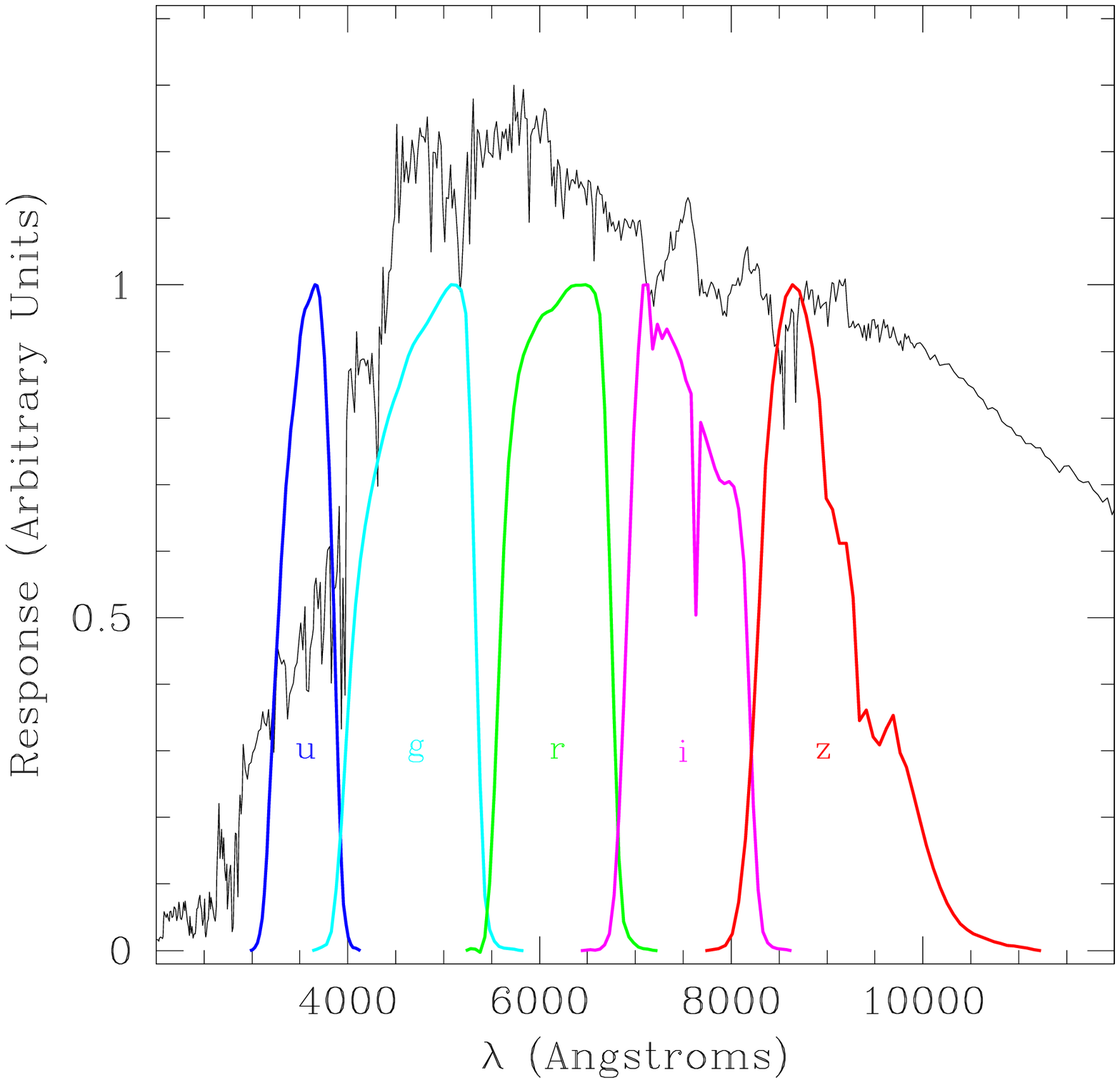}
\caption{\label{response_sdss} Estimated response for all five bands
in the SDSS (accounting for the atmosphere, the mirrors, the filters,
and the CCDs), as a function of observed wavelength, as measured by
Mamoru Doi. The predicted spectrum $f(\lambda)$ for a 4 Gyr old
instantaneous burst using the models of \citet{bruzual93a} (and
observed at $z=0$) is shown for reference.}
\end{figure}

The response functions $R(\lambda)$ of the filter-CCD combinations
have been measured using a monochrometer by Mamoru Doi.  Using a model
for the atmospheric transmission and the reflectivity of the primary
and secondary mirrors, one can then model the response of the entire
system. For each bandpass in the SDSS, there are six different CCDs;
it has been shown that the differences are small between these CCD and
filter responses for each bandpass.  The resulting set of filter
curves is shown in Figure \ref{response_sdss}, in comparison to a
model of a galaxy spectral energy distribution observed at $z=0$ (a 4
Gyr old instantaneous burst from the models of \citealt{bruzual93a}).

The results of the spectroscopic observations are treated as follows.
We extract one-dimensional spectra from the two-dimension images using
a pipeline ({\tt specBS v4\_8}) created specifically for the SDSS
instrumentation (\citealt{schlegel02a}), which also fits for the
redshift of each spectrum. The official SDSS redshifts are obtained
from a different pipeline (\citealt{subbarao02a}). The two independent
versions provide a consistency check on the redshift determination,
and for galaxies they agree on more than 99\% of the
objects.\footnote{The disagreement is more significant for objects
with unusual spectra, such as certain types of stars and QSOs.}

We use two types of magnitudes determined by the SDSS. First, the SDSS
Petrosian magnitudes, a modified form of the magnitude described by
\citet{petrosian76a}, as described in \citet{strauss02a}.  Petrosian
magnitudes are circular aperture magnitudes with an aperture size
which is determined by the shape of the radial profile (not its
amplitude). The resulting apertures are empirically nearly constant in
metric size as a function of redshift; in addition, all the bands use
the same aperture, so the measured SED corresponds (to within the
effects of seeing) to the SED of an identifiable region of the
galaxy. However, for faint objects, the Petrosian magnitudes tend to
become noisy. Thus, for galaxies with $z>0.25$ we instead use the
higher signal-to-noise ``model magnitudes.'' Model magnitudes are
calculated in all bands using a single weighted aperture convolved
with the point spread function; thus, the measured colors again
correspond to an identifiable region of the galaxy, though a different
one than would be measured by the Petrosian magnitude. The weighted
aperture is the better fitting model (pure exponential or pure de
Vaucouleurs) to the galaxy image in the $r$-band. Thus, the model
magnitudes weight the centers of galaxies more strongly than do
Petrosian magnitudes. In \citet{stoughton02a} we are explicitly warned
not to mix Petrosian and model magnitudes in a single analysis, since
they measure galaxies in very different ways. Nevertheless, for the
purposes of contraining galaxy template SEDs, it is perfectly
acceptable to use {\it any} well-defined part of any galaxy.  For our
purposes, it is more important to have high signal-to-noise
measurements of galaxy colors than to measure exactly the same regions
of galaxies at low and high redshift.  We extinction-correct both
types of magnitudes using the dust maps of \citet{schlegel98a}.

\subsection{Fitting SEDs to SDSS Data}

For the purposes of using the SDSS data to constrain our template set,
we take a subsample of the data consisting of around 30,000 objects in
the range $0.0<z<0.5$. The sample includes both the main sample and
the LRGs, and is designed such that there is an approximately even
distribution over redshift within our range of redshifts. In addition,
we add results from galaxies in several spectroscopic plates
(totalling about 1,000 objects) which were selected by a photometric
redshift algorithm (Csabai {\it et al.}, in preparation) to be at
around $z\sim 0.3$--$0.4$, and subsequently observed
spectroscopically. These objects are invaluable for tying down the
blue end of the templates.  Finally, we exclude galaxies in the
redshift range $0.28 <z<0.32$ from the fit for the templates, for
reasons which we will explain more fully below (nevertheless, we still
can and do use the resulting templates to analyze galaxies in this
range).

Some of the objects have missing or poorly constrained data. For
example, the $u$- or $z$-band fluxes for some objects are swamped by
the photon noise of the sky. We identify such cases as magnitude
errors greater than 2.0 or magnitudes fainter than 24.0 in any
band. We ignore these objects entirely when fitting for the
templates. 

In addition, the photometric errors for most objects in the
spectroscopic galaxy sample of the SDSS are not dominated by the
estimated errors on the photometry listed in the catalog. Instead, the
errors are dominated by local calibration errors and other systematic
effects, which are poorly known. To account for these errors, we add
extra error terms in quadrature with the errors listed in the catalog
(0.05 mags, 0.02 mags, 0.02 mags, 0.02 mags, and 0.03 mags for
$ugriz$, respectively). The choice of these values is based on a
qualitative sense of the photometric errors present in the data. Not
accounting for this extra source of errors can cause the fits to be
ill-behaved.

For the SED space (our $\vv{v}_i(\lambda)$), we use the subspace
defined by ten Bruzual-Charlot instantaneous burst models with ages
ranging from $3 \times 10^7$ to $2\times 10^{10}$ years, five with
metallicity $Z=0.02$ and five with $Z=0.004$, all assuming a Salpeter
Initial Mass Function. Since we cannot recover information about the
SED below a certain wavelength resolution, we smooth most of the
wavelength regime of the templates using a Gaussian with a standard
deviation of $\sigma = 300$\AA; for the region of the spectrum which
contains the sharpest gradients, around 4000\AA, we smooth only with
$\sigma = 150$\AA. None of our results change dramatically if we vary
our smoothing procedure or if we add reddened templates to our allowed
SED space.

We choose to fit for $N_t = 4$ eigentemplates, the maximum one can use
and still allow freedom to fit for the templates themselves.  Simply
using five templates (which obviously reproduces all the magnitudes
exactly) tends to yield unphysical trends of galaxy SED versus
redshift ({\it cf.}  Figure \ref{main_colors_plot} below). Using three
templates does nearly as well as four templates in the sense that the
resulting templates reproduce the $griz$ magnitudes nearly as
well. However, the fourth template is necessary to recover the
$u$-band flux to better than about 15\%. In addition, since one of the
applications of these SED determinations is the distribution of galaxy
colors in fixed frame magnitudes, we don't want to artificially reduce
the dimensionality of the color space to only two.

One more choice needs to be made, the wavelength regime over which to
orthogonalize the templates and over which to calculate the flux. We
choose the range defined by $\lambda_{\mathrm{min}}=3500$\AA\ and
$\lambda_{\mathrm{max}}=7500$\AA, since this rest frame range is
observed for almost all galaxies in the sample. We refer here and in
other papers to the flux and luminosity in this range as the ``visual
flux'' $f_v$ and the ``visual luminosity'' $L_v$.

Once we have finished fitting for the templates, we still want to
determine a best-fit SED for each object.  Thus, after the templates
have been determined, we fit for the coefficients $a_i$ for each
galaxy, including those excluded from the template-fitting procedure
because of large errors. In order to accommodate the presence of
errors, we constrain all the SED fits to all galaxies using the method
described in Section \ref{direct}. As a constraint, we use the
covariance matrix and mean value (essentially zero, by design) of the
coefficients of the well-measured set of galaxies used to determine
the templates; we multiply all the elements of the covariance matrix
by a factor of four to weaken our constraint (which is appropriate,
since the actual distribution of coefficients is not well described by
a single gaussian).  Thus, we account for any missing information by
simply requiring that the object SED have ``reasonable''
properties. The constraint does not affect the results for galaxies
which have well-measured magnitudes in all bands.

\subsection{The $g$/$r$ Gap}
\label{grgap}

In Figure \ref{response_sdss}, there is clearly a gap between the $g$
and $r$ bands. Linear fits to the data allow considerable freedom in
the resulting reconstructed SEDs. As it happens, one of the directions
in our best fit coefficient space $a_i/a_0$ corresponds to a large
spike at around 4000 \AA; that this direction is important is not
surprising, since one of the most variable quantities of galaxy SEDs
is the size of the 4000 \AA\ break. However, the gap between the $g$
and $r$ bands corresponds to the 4000 \AA\ break at around
$z=0.3$. The existence of this spike in one direction in our
four-dimensional space means that galaxies at $z=0.3$ can vary along
this direction in order to better fit $u$, $i$, and $z$, with little
effect on the $g-r$ color of the fit. This results in unphysical SED
fits to galaxies near $z=0.3$.  One really only needs to worry about
this effect in the SDSS when one is dealing with the LRGs, for
galaxies in the range about $0.27 < z < 0.33$. The Main Sample results
are not significantly affected at all (since few Main Sample galaxies
are at $z>0.25$.  Furthermore, one can still observe a galaxy at
$z=0.1$ and reliably infer what it would look like at $z=0.3$; it is
only the reverse process which is difficult.

\begin{figure}
\figurenum{\fignum}
\plotone{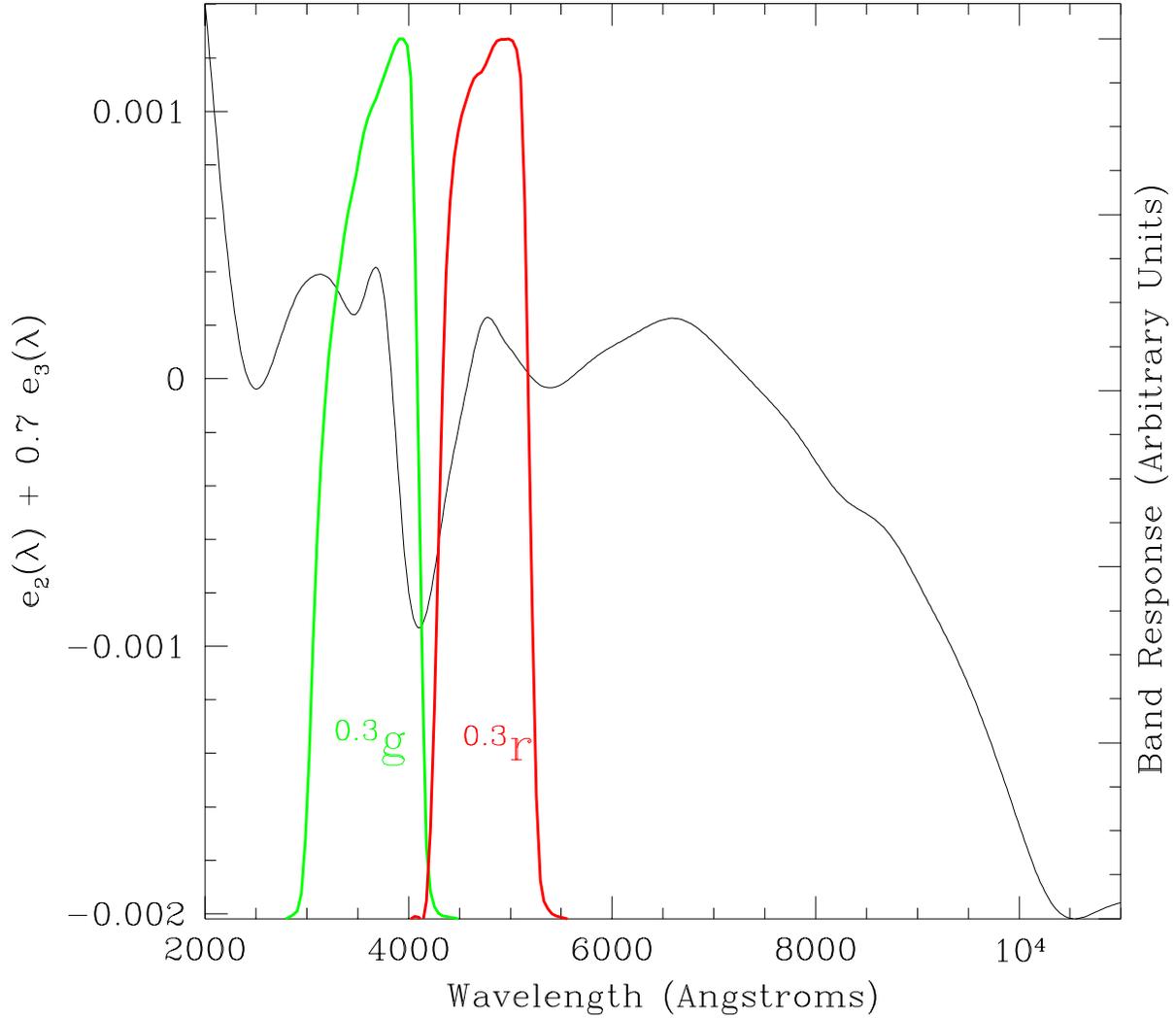}
\caption{\label{spur} The spectrum corresponding to the direction of
the spur in the upper right panel of Figure
\ref{k_coeffdist_plot}. Note the strong feature near 4000
\AA. Overplotted are the bandpasses for $\band{0.3}{g}$ and
$\band{0.3}{r}$. The strong features fall in the gap between the
bandpasses. Thus, in a linear fit to a galaxy at $z=0.3$, this
component can be used to fit the observed magnitudes without being
constrained to have reasonable behavior around 4000 \AA. }
\end{figure}

To demonstrate this degeneracy, Figure \ref{spur} shows the spectrum
corresponding to the direction represented by the vector $ \vv{e}_2 +
0.7 \vv{e}_3$, for our best fit templates. We overplot the
\band{0.3}{g} and \band{0.3}{r} bands. This direction has a strong
peak in the gap between the two bands at $z=0.3$.  This difficulty is
why, in Section \ref{data}, we exclude the regime around $z\sim 0.3$;
otherwise the templates became distracted by the degeneracies in this
range.

We have explored using a larger set of input spectra, by including
reddened versions of all our templates, to test whether a more
complete space of SEDs would cause our fit to choose a more realistic
subspace. In addition, we have tried removing the increased resolution
of the template SEDs around the 4000 \AA\ break. Neither of these
tests yielded better results.

Our opinion is that the ideal regularization which would solve this
problem would be to minimize $\chi^2$ under the constraint that none
of the original Bruzual-Charlot spectra defining our SED space
contribute negatively to the SED fit. Such a solution would almost
certainly have better properties than our linear approach here: it
would almost certainly avoid the degeneracies, it would yield
reasonable fit SED over a larger wavelength range, it would associate
a reasonable star-formation history with each object (which could be
used to evolution-correct the magnitudes).

However, we have not tackled this approach yet. As it happens, the
direct constraint on the coefficients described in the previous
section is sufficient to suppress the $g$/$r$ gap problem 
for many purposes, as long as one $K$-corrects the magnitudes to
$z=0.3$. In this way, one minimizes the $K$-corrections for the
galaxies which one is least certain of.

\subsection{Results}

The top two panels of Figure \ref{k_coeffdist_plot} show the pairwise
joint distributions of $a_1/a_0$, $a_2/a_0$, and $a_3/a_0$ for a
random subset of about 10,000 of the galaxies in the SDSS sample (not
only the ones which we used to fit the templates). In the bottom
panel, three spectra taken from a one-dimensional sequence along the
galaxy locus are shown, showing that the spectra become progressively
bluer along that sequence. Note that constraints on the SED become
poor at the bluest and reddest edges of this diagram (the peak of the
$z$-band response is at 8700~\AA rest frame for a galaxy observed at
$z=0$). Therefore, the odd behavior near the edges should not be taken
too seriously. In addition, note the small spur extending from lower
left to upper right through the right edge of the
($a_3/a_0$)-($a_2/a_0$) plane. Many of the objects in this spur are at
around $z=0.3$, and this spur is the result of the degeneracy due to
the $g$/$r$ gap.

\begin{figure}
\figurenum{\fignum}
\plotone{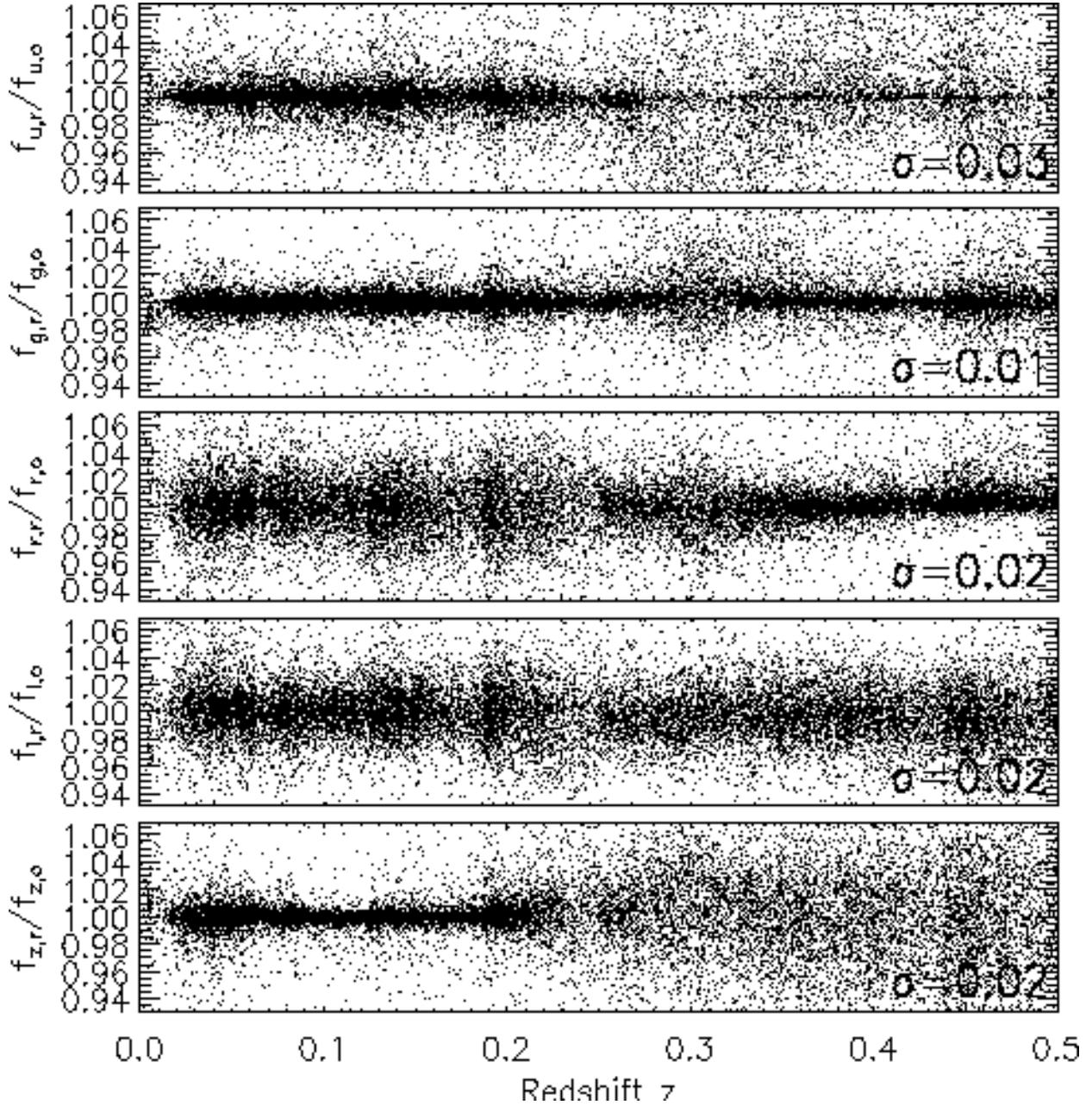}
\caption{\label{k_model_plot} Reconstructed galaxy fluxes relative to
the observed galaxy fluxes, for all five SDSS bands, shown for a
random subsample consisting of around 10,000 of the SDSS galaxies. The
residuals are shown against redshift.  There is no systematic trend
with redshift in any band. The 5-$\sigma$ clipped estimate of the
scatter around the observed fluxes is listed for each band. In $u$,
$g$, $r$, and $i$ the scatter is consistent with the expected
photometric errors in the survey at all redshifts. At high redshift
the scatter in $z$ becomes large, most likely due to increasing
photometric errors. }
\end{figure}

These reconstructed spectra do an excellent job of reproducing the
observed galaxy fluxes. Figure \ref{k_model_plot} shows the
differences between the observed and reconstructed fluxes as a
function of redshift. There are no systematic trends with redshift,
and the standard deviations of the differences between the observed
and the reconstructed fluxes (shown on the right bottom corner of each
panel) are of order the photometric errors in the sample. In
particular, for our 30,000 galaxies (and thus approximately 30,000
degrees of freedom) we find $\chi^2 \sim 55,000$, indicating that our
fit is reasonable. 

\begin{figure}
\figurenum{\fignum}
\plotone{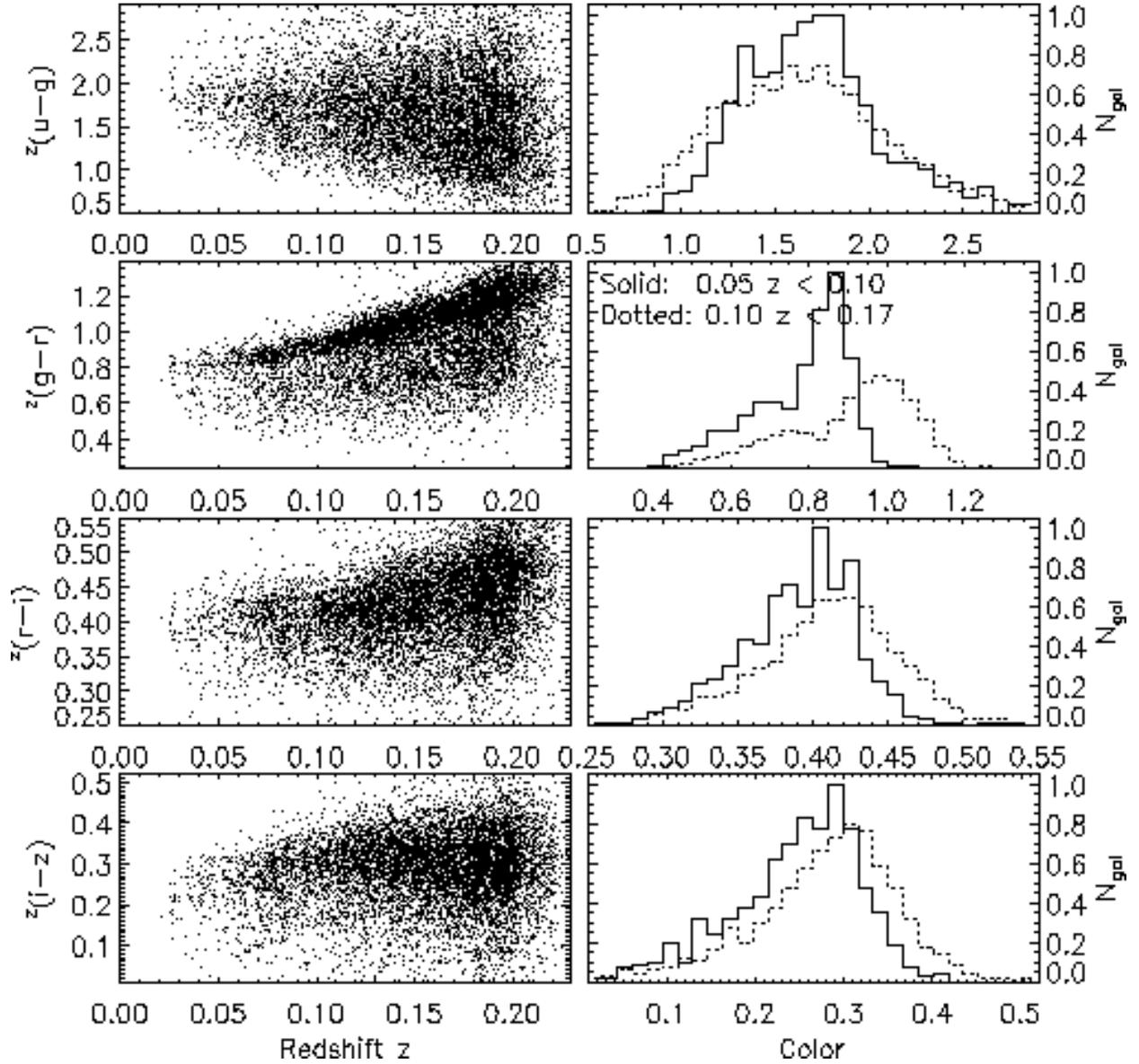}
\caption{\label{main_colors_plot.z} Color distributions in the
observed frame for SDSS Main Sample galaxies in the luminosity range
$-21.5<M_{\band{0.1}{r}}<-21.2$. This sample is complete (that is,
volume limited) for $0.05<z<0.17$. Left panels show the colors as a
function of redshift. Right panel shows the distributions of each
color at high and low redshift within the volume-limited
subsample. The observed colors clearly depend strongly on redshift.}
\end{figure}

An important test of the consistency of the fits is to check that for
a fixed type of galaxy, the distribution of the fixed frame colors
depends only weakly on redshift. Because the galaxies shown in Figures
\ref{k_coeffdist_plot} and \ref{k_model_plot} are inhomogeneously
selected --- some are Main Sample galaxies, some are LRGs, and some
are selected from the photometric redshift plates --- we will split
our sample into two well-defined sets of objects to perform this
test. First, we choose a set of Main Sample galaxies in the luminosity
range $-21.5 < M_{\band{0.1}{r}} < -21.2$. Figure
\ref{main_colors_plot.z} shows the observed-frame colors of all the
galaxies as a function of redshift.  The right-hand panels show the
distribution for redshifts $0.05<z<0.10$ (solid histogram) and
$0.10<z<0.17$ (dotted histogram). A trend exists in all colors, most
strongly in $\band{z}{(g-r)}$. Figure \ref{main_colors_plot} shows the
$K$-corrected colors $\band{0.1}{(u-g)}$, $\band{0.1}{(g-r)}$,
$\band{0.1}{(r-i)}$, and $\band{0.1}{(i-z)}$ in the same manner.  The
plots show a general consistency in the fixed frame color distribution
of these objects with redshift.  Small changes are discernible in the
distributions, mostly attributable to the increased errors at higher
redshift. Note that for $z<0.1$ the $\band{0.1}{(u-g)}$ color depends
on an extrapolation of the SED in the blue, while for $z>0.1$ the
$\band{0.1}{(i-z)}$ color depends on an extrapolation of the SED in
the red.

\begin{figure}
\figurenum{\fignum}
\plotone{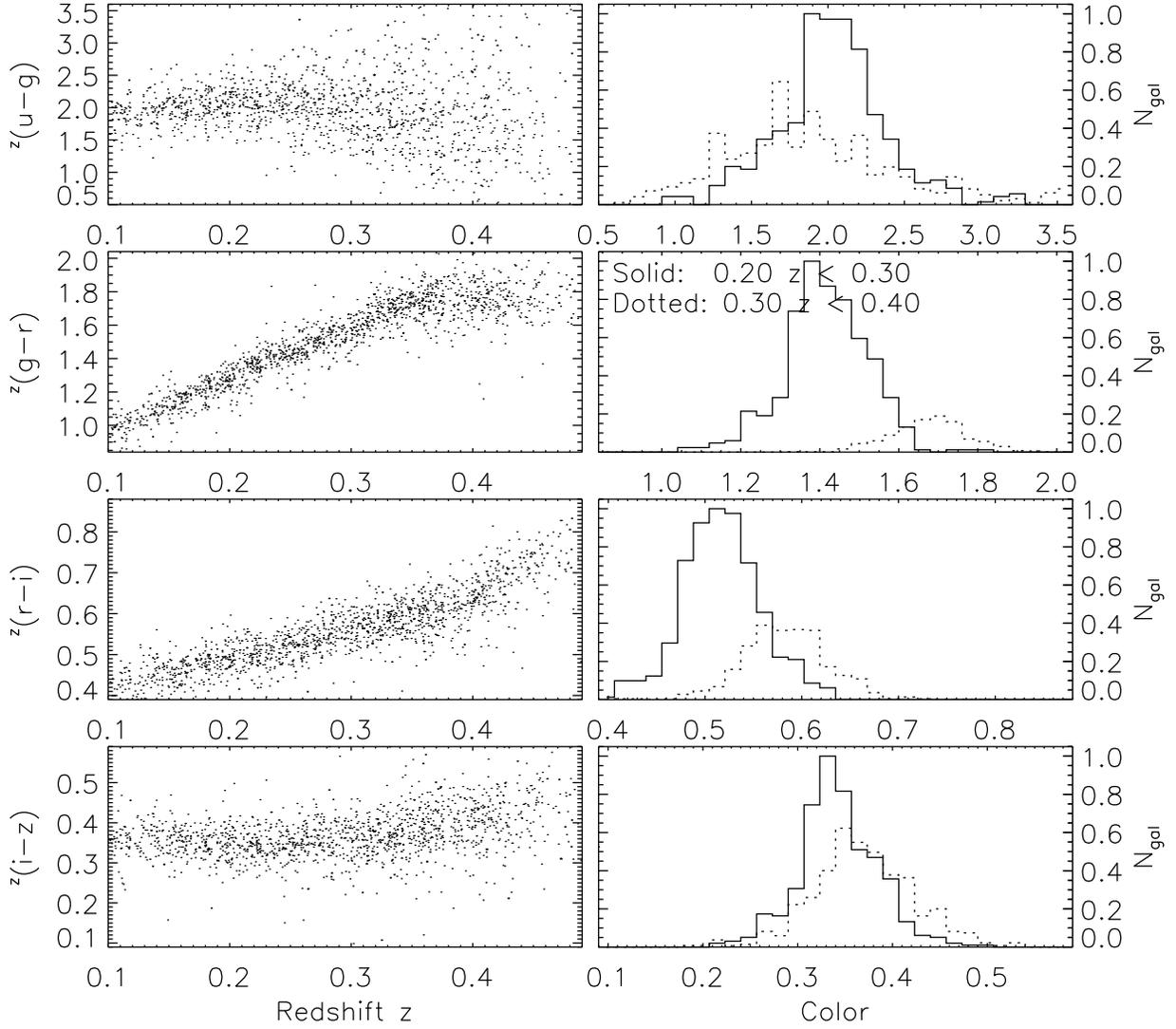}
\caption{\label{lrg_colors_plot.z} Similar to Figure
\ref{main_colors_plot.z}, now showing LRG galaxies (Cut I) in the
luminosity range $-22.8<M_{\band{0.3}{r}}<-22.5$. Again, there is a
strong dependence on redshift.}
\end{figure}

\begin{figure}
\figurenum{\fignum}
\plotone{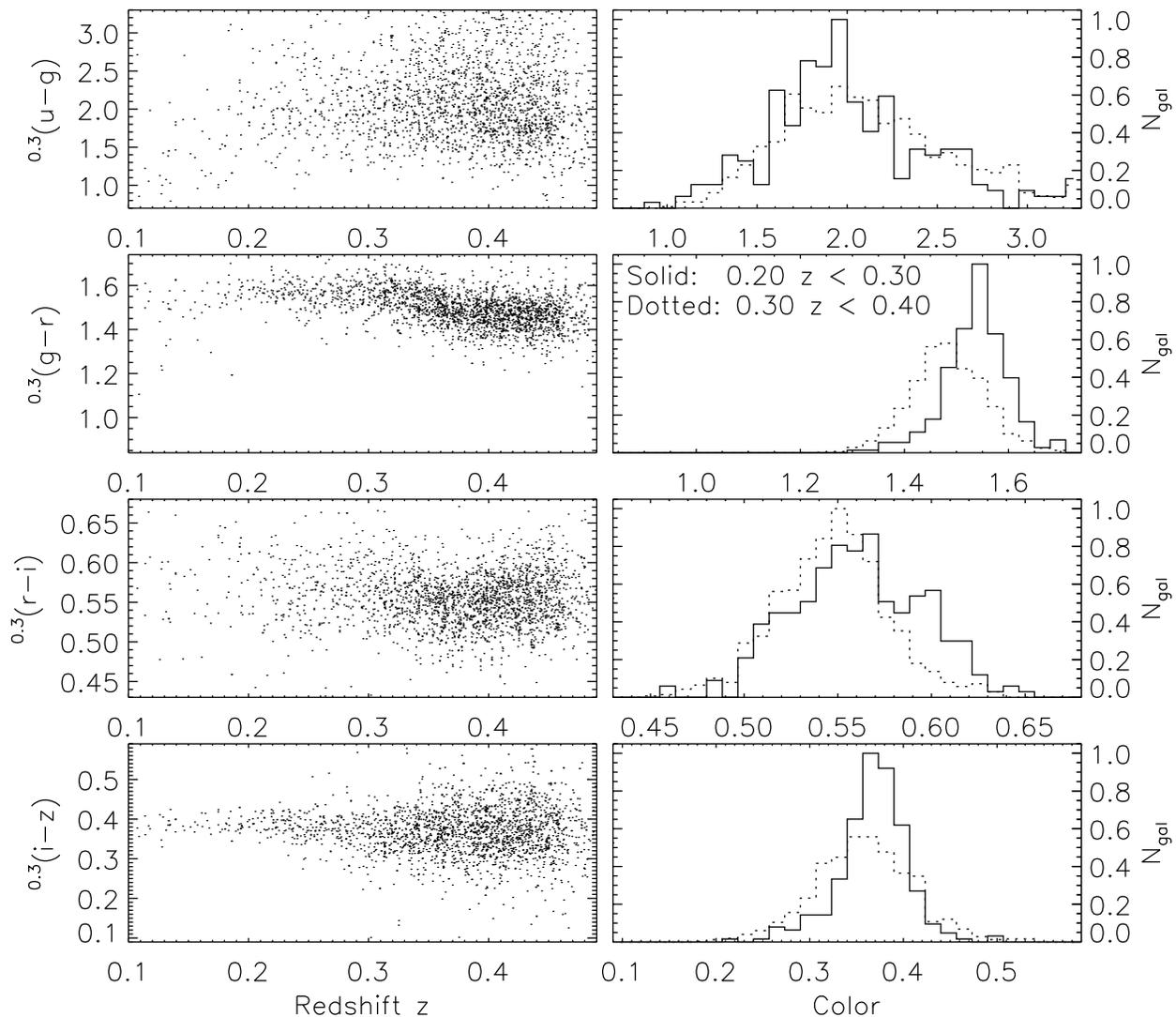}
\caption{\label{lrg_colors_plot} Same as Figure
\ref{lrg_colors_plot.z}, now $K$-correcting the LRG galaxies to
$z=0.3$.  The redshift dependence is greatly reduced for the LRGs in
comparison to Figure \ref{lrg_colors_plot.z}; on the other hand, there
are distinct trends of restframe color with redshift.  In
$\band{0.3}{(g-r)}$ an overall trend is apparent; LRGs at $z=0.4$ are
about 0.1 magnitudes bluer than LRGs at $z=0.2$. A change of this
magnitude is attributable to passive galaxy evolution, though
considerably more work needs to be done to show that this is
occurring. In $\band{0.3}{(r-i)}$, a blueward shift also occurs,
though at a much smaller level.}
\end{figure}

Second, we choose a set of Cut I LRGs (see \citealt{eisenstein01a} for
details) with $-22.8 < M_{\band{0.3}{r}} < -22.5$. Figure
\ref{lrg_colors_plot.z} shows the observed colors as a function of
redshift for these objects. Again, there is a strong redshift
dependence. Figure \ref{lrg_colors_plot} shows the colors
$K$-corrected to $z=0.3$, which appear approximately constant with
redshift.
$\band{0.3}{(g-r)}$ (which is very similar to $\band{0.0}{(u-g)}$)
experiences a blueward shift of about 0.1 magnitudes between $z=0.1$
and $z=0.45$, which is attributable to passive evolution, though it
could also be due to the LRG selection procedure.

In short, these fits to galaxy SEDs provide estimates which reproduce
the galaxy photometry nearly to the level of the errors in the
photometry itself, seem physically reasonable, and are consistent over
the range of redshifts we consider ($0.0 < z < 0.5$). 

\subsection{$K$-corrections in the SDSS}

\begin{figure}
\figurenum{\fignum}
\plotone{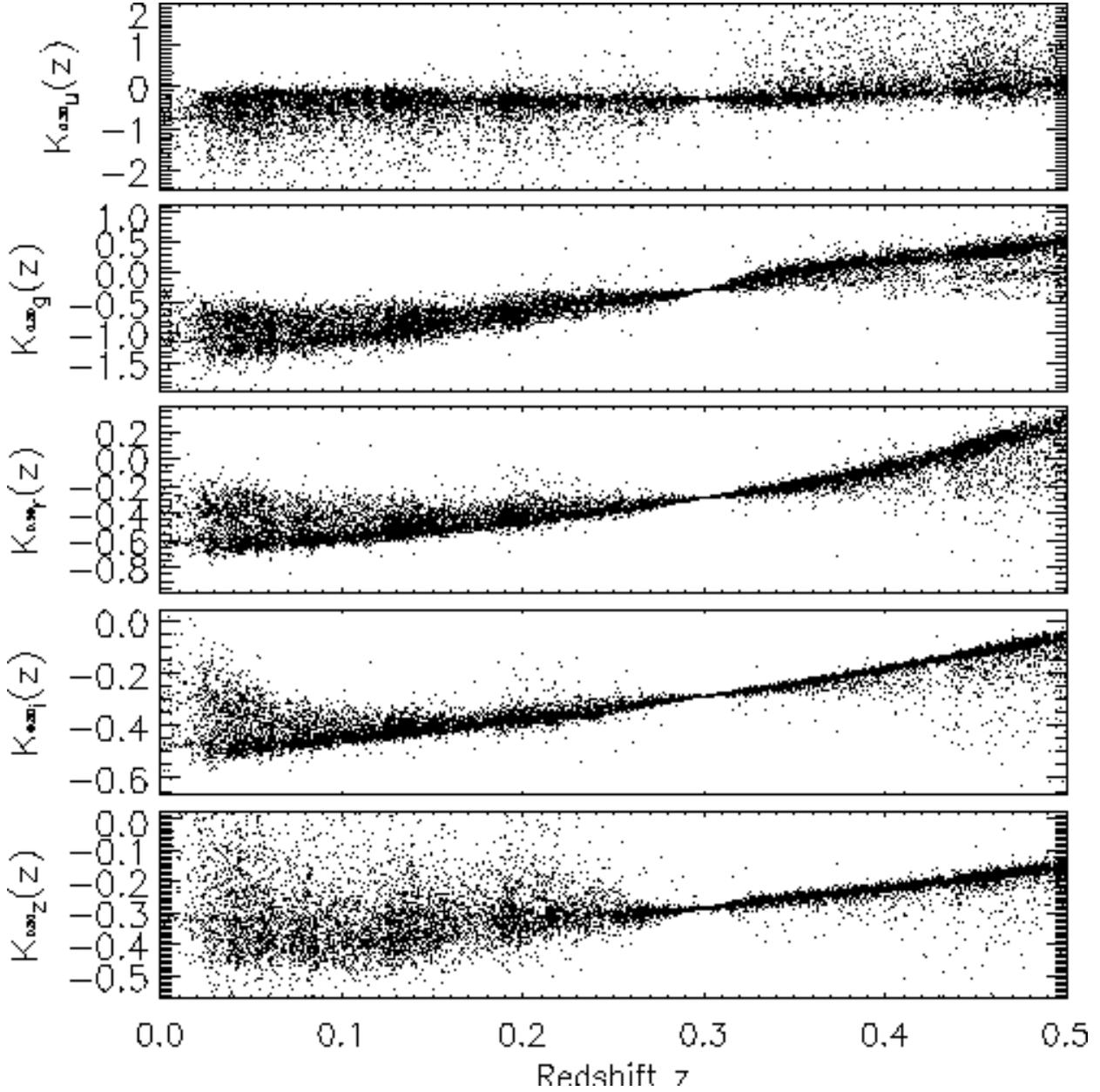}
\caption{\label{k_kcorrect_plot} $K$-corrections to $z=0.3$ as a
function of redshift in all five bands for a random subsample
consisting around 10,000 of the SDSS galaxies.  The range of
$K$-corrections at each redshift reflects the range of galaxy types at
each redshift.  The $K$-corrections are largest, and therefore the
most uncertain, for the \band{0.3}{u} and \band{0.3}{g} bands. While
we show the $K$-corrections for \band{0.3}{u} at $z<0.3$ and for
\band{0.3}{z} at $z>0.3$, and indeed these $K$-corrections are fairly
well-behaved, we do not recommend using these extrapolated results for
scientific purposes. }
\end{figure}

We show, in Figure \ref{k_kcorrect_plot}, the resulting $K$-corrections to
$z=0.3$ inferred from this method, for all five SDSS bands. Note that
the $K$-corrections are largest (and thus most uncertain) in
$\band{0.3}{u}$ and $\band{0.3}{g}$. Remember that the $K$-corrections
to $\band{0.3}{u}$ are extrapolations for $z<0.3$ and that the
$K$-corrections to $\band{0.3}{z}$ are extrapolations for $z>0.3$, so
those results should not be taken too seriously. (Although the
$K$-corrections given are not {\it too} unreasonable).

An important test of our method is to synthesize broad band photometry
from spectra, and then try to recover the $K$-corrections in a case
where we know the spectrum completely. For this purpose, we use as
example spectra the spectra of galaxies in the SDSS.  Since we cannot
synthesize the observed $u$ and $z$ bands from these spectra, we
simply base them on the synthesized $g$ and $i$ band magnitudes plus
the actual $u-g$ and $z-i$ colors. (This procedure is unlikely to make
the photometric estimations of the $K$-corrections {\it
better}). Finally, we added 2\% random errors to all of the
measurements. Figure \ref{k_speck_plot.fitfib.0.1} compares the
$K$-corrections from the synthesized photometry to that calculated
from the spectrum itself:
\begin{equation}
\Delta K = K_{\mathrm{spec}} - K_{\mathrm{photo}}
\end{equation}
Since all of the photometry used in this test is actually synthesized
from the spectra, this is {\it not} a comparison of photometric and
spectroscopic $K$-corrections; it is {\it only} a test of how well our
method recovers $K$-corrections. The agreement in FIgure
\ref{k_speck_plot.fitfib.0.1} is very good, suggesting that our method
can indeed recover the correct $K$-corrections based only on broad
band magnitudes.

\begin{figure}
\figurenum{\fignum}
\plotone{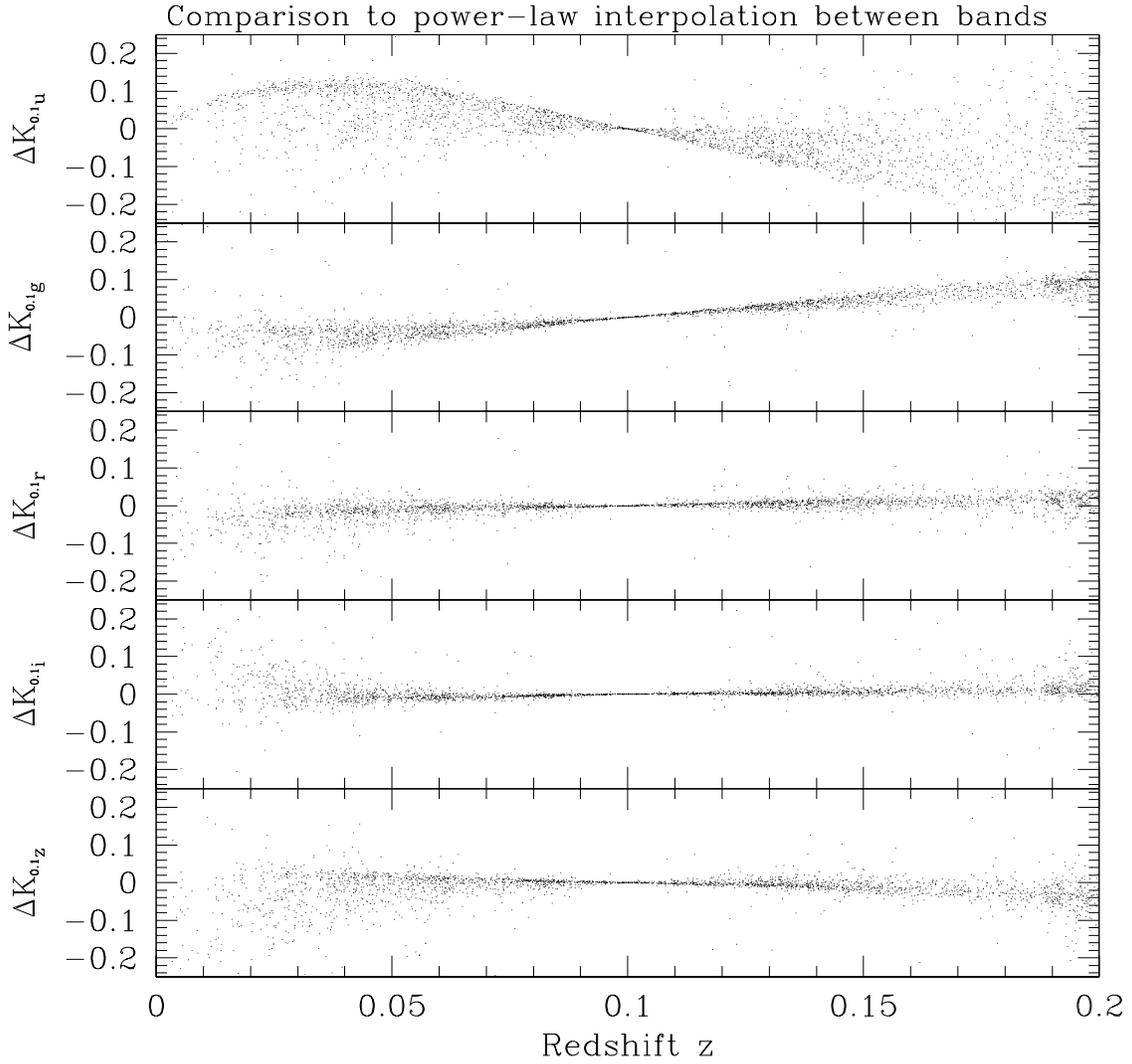}
\caption{\label{compareci} Difference in the $K$-corrections to
$z=0.1$ in each band between the method used in Figure
\ref{k_kcorrect_plot} and the method of simply interpolating between
adjacent bands fitting a power-law SED. The differences are small in
\band{0.1}{r}, \band{0.1}{i}, and \band{0.1}{z}, where galaxy SEDs
have simple shapes. There are large systematic differences in
\band{0.1}{u} and \band{0.1}{g}, for which the 4000 \AA\ break is
important in the spectral templates used. }
\end{figure}

\begin{figure}
\figurenum{\fignum}
\plotone{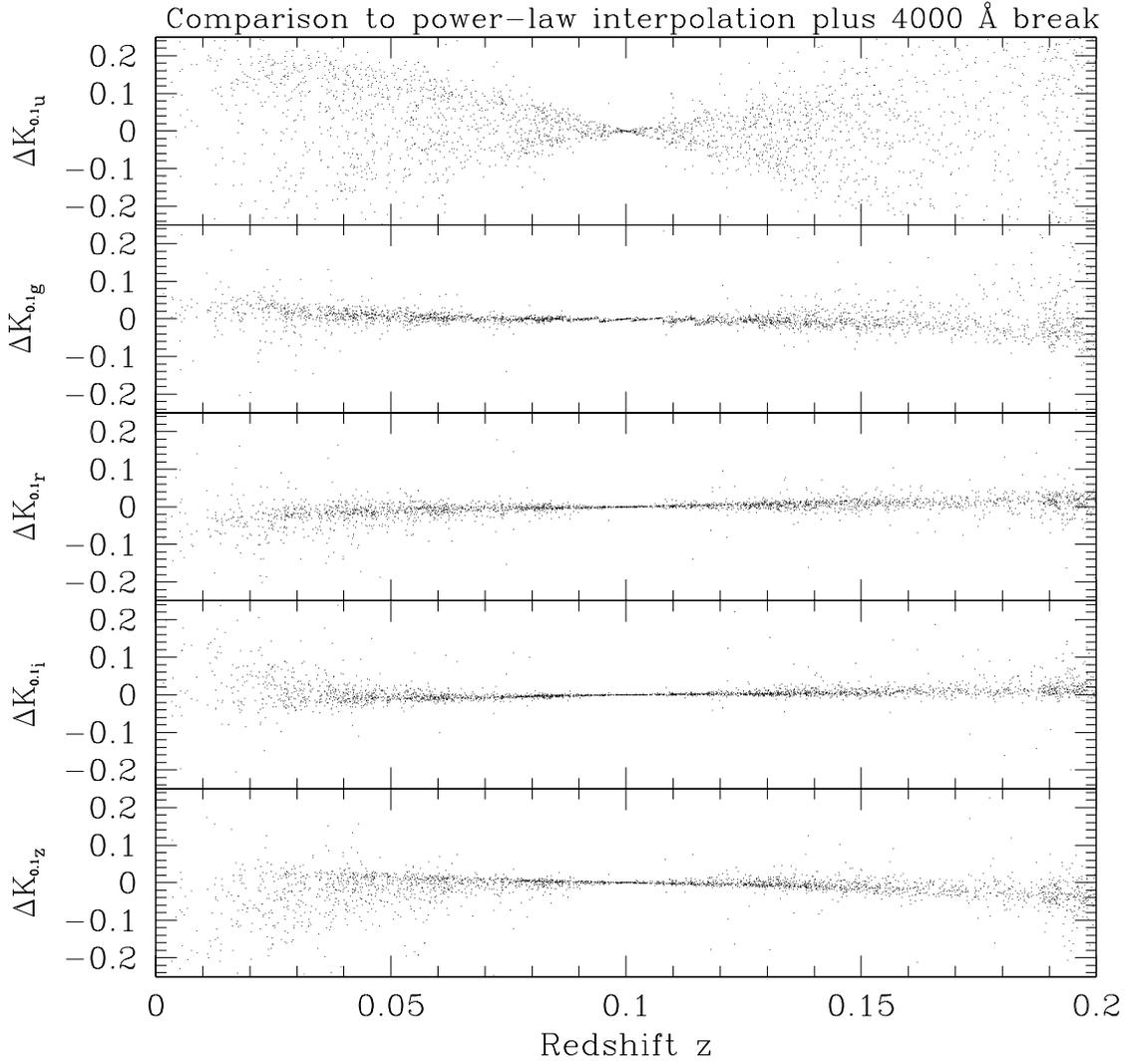}
\caption{\label{comparecibreak} Same as Figure \ref{compareci},
now comparing to a method of interpolating the bandpasses using
power-laws, and fitting for the 4000 \AA\ break. The systematic trends
in \band{0.1}{u} and \band{0.1}{g} are gone (though there is
considerable scatter in \band{0.1}{u}). }
\end{figure}

To test how robust these $K$-correction results are to our model
assumptions, we compare them to other methods of calculating fixed
frame galaxy magnitudes.  An extremely simple method is to calculate
the flux in any desired bandpass by fitting a power-law slope and
amplitude to the fluxes in the two adjacent bandpasses (extrapolating
when necessary). Figure \ref{compareci} shows the differences in
the $K$-corrections inferred from this method and those inferred from
the method of Figure \ref{k_kcorrect_plot}, as a function of
redshift. $r$, $i$, and $z$ are all reasonably similar in either
method; $u$ and $g$, however, have distinct trends with redshift,
mostly due to the non-power law nature of the template galaxy spectra
(and probably actual galaxy spectra) in this regime. To show this
fact, we perform a similar power-law fit, only this time including a
break in the spectrum at 4000 \AA. We use the $u$-$g$ color to fit the
break, assuming that the slope blueward of 4000 \AA\ is
$f(\lambda)\propto \lambda^{2}$.  Figure \ref{comparecibreak} shows
the results of this fit; the redshift trend in $g$ is greatly reduced,
as is the trend in $u$, but a large amount of scatter remains in the
$u$ band. This results from the fact that you can {\it either} fit the
slope of the SED below the 4000 \AA\ break {\it or} the size of the
4000 \AA\ break itself; it is not possible with this data to constrain
both in an individual spectrum, which is a limitation of our
determination of fixed frame $u$-band magnitudes.

Finally, it is possible to use the galaxy spectra obtained with the
spectrograph to estimate the $K$-correction for each object.  However,
this procedure requires trusting the spectrophotometry over a wide
wavelength range. In addition, the region of the galaxy which the
spectra cover is not completely well defined. The small scale
wavelength features are set by the fiber aperture ($3''$ in diameter)
while large scale wavelength features are constrained by a short
``smear exposure'' which covers a larger area. The smear exposure
corrections are clearly the correct approach for point sources, but
for galaxies it causes the small scale and large scale features of the
spectrum to be determined by somewhat different regions of the galaxy.

\begin{figure}
\figurenum{\fignum}
\plotone{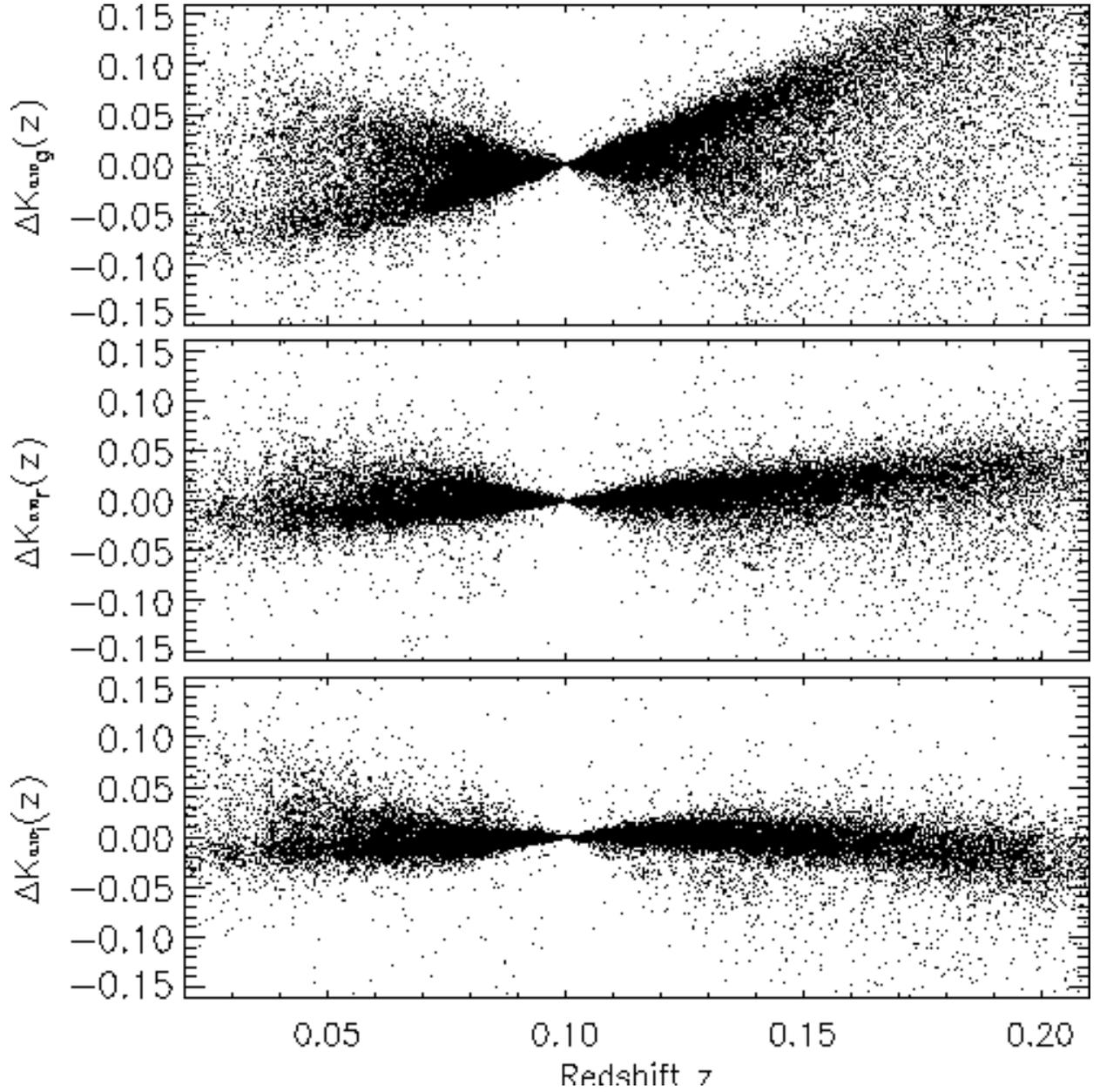}
\caption{\label{k_speck_plot.0.1} Difference between the
$K$-corrections to $z=0.1$ determined from the spectroscopy and those
determined from the analysis of the broad-band Petrosian magnitudes, for
Main Sample galaxies.}
\end{figure}

Nevertheless, in Figure \ref{k_speck_plot.0.1}, we compare the
$K$-corrections to $z=0.1$ of Main Sample galaxies calculated based on
the spectra to those of Figure \ref{k_kcorrect_plot}, finding that
they are quite similar. There is a trend with redshift in the
$g$-band, which is due to the fact that on average the $g-r$ color is
about 0.1 magnitudes redder in the fiber aperture than in the
Petrosian aperture. The implication is that the $K$-corrections are
stronger for the aperture which corresponds to the spectrum than for
the Petrosian aperture. This indicates that for work which requires
precision, it is best to avoid using the $K$-corrections determined
from the spectra. (Note that in the second paragraph of this section
we used the SDSS spectra as {\it example} spectra to validate the
method as applied to synthesized photometry; in this paragraph we have
evaluated whether to rely on fiber spectra to calculate
$K$-corrections for the Petrosian photometry).

\begin{figure}
\figurenum{\fignum}
\plotone{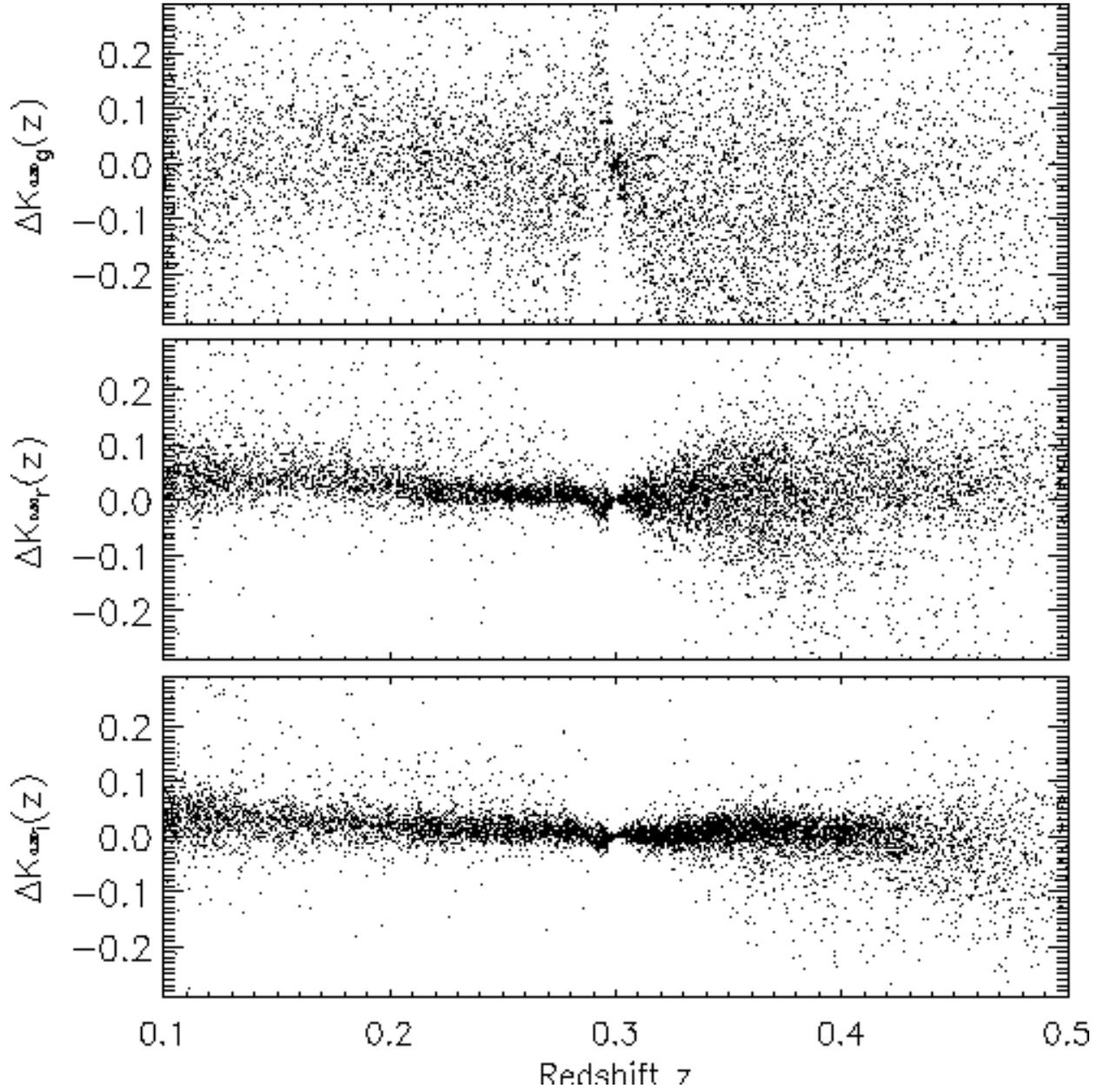}
\caption{\label{k_speck_plot.0.3} Difference between the
$K$-corrections to $z=0.3$ determined from the spectroscopy and those
determined from the analysis of the broad-band model magnitudes, for
LRGs.}
\end{figure}

For completeness, Figure \ref{k_speck_plot.0.3} compares the
spectroscopic and photometric $K$-corrections to $z=0.3$ of LRGs. The
\band{0.3}{r} and \band{0.3}{i} bands agree very well. However, there
is considerable scatter in the \band{0.3}{g} band (nearly equivalent
to the \band{0.0}{u} band), at around 20\%. Because the LRGs tend to
have small color gradients, the issues of fiber sizes and smear
exposures are not important in this context.

\section{Conclusions and Future Work}
\label{conclusions}

We have presented a method and an implementation for estimating galaxy
SEDs for the purpose of calculating fixed frame galaxy magnitudes over
a range of redshifts. We have demonstrated that it gives sensible and
consistent results. We will be using this method in future papers
which will describe the joint distribution of luminosities and colors
of galaxies, as well as the evolution of the luminosity function of
galaxies. Furthermore, we plan to incorporate observations of objects
in bands other than the SDSS bands to further describe the nature of
galaxy SEDs.

\begin{figure}
\figurenum{\fignum}
\plotone{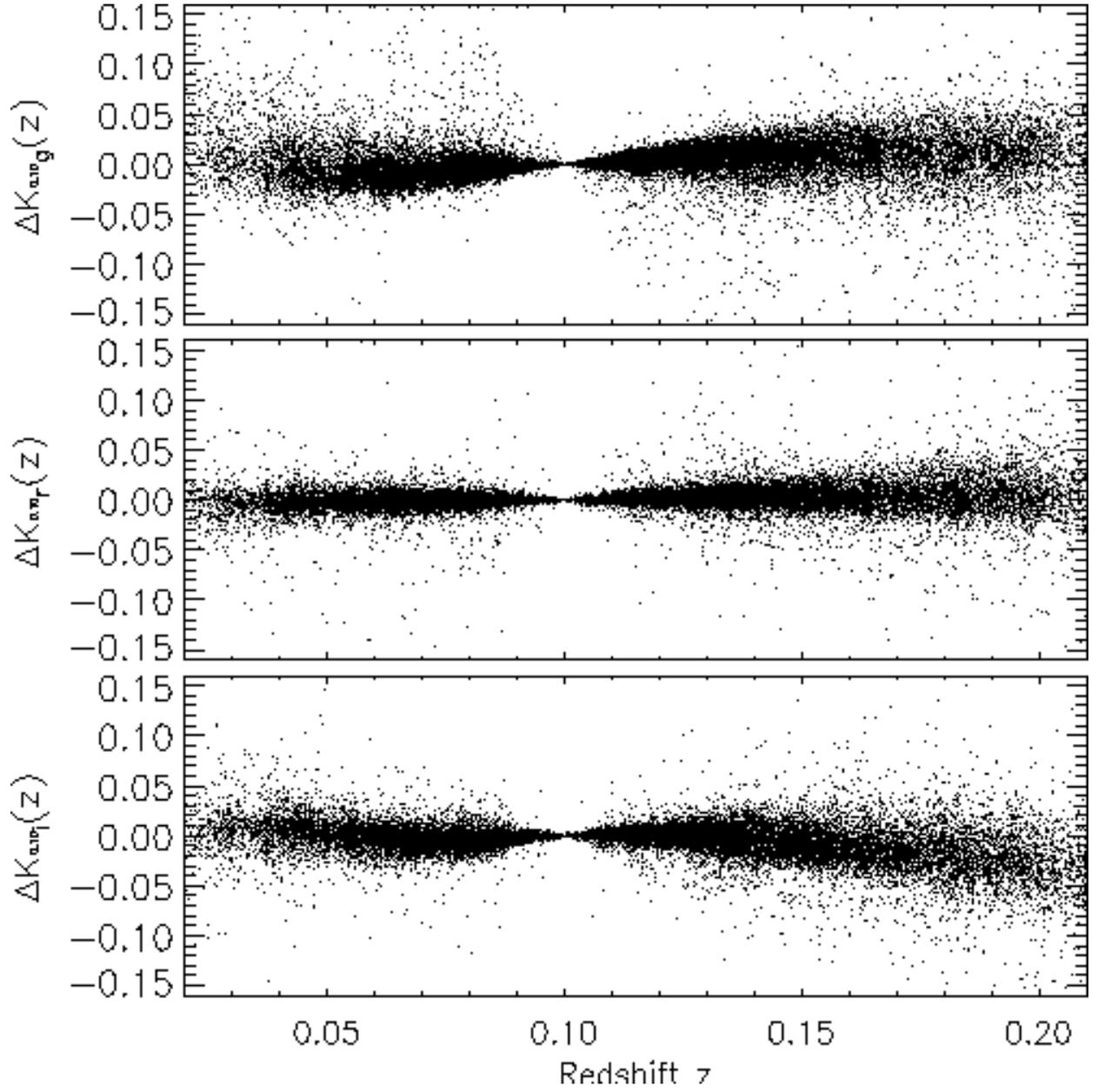}
\caption{\label{k_speck_plot.fitfib.0.1} Difference between the
$K$-corrections to $z=0.1$ determined from the spectroscopy and those
determined from the analysis of broad-band magnitudes {\it
synthesized} from the same spectra.}
\end{figure}

We have not discussed in detail the question of error analysis. There
is no answer to this question which is simple, general, {\it and}
realistic. The photometric errors can of course be propagated to the
errors on the reconstructed magnitudes, but these estimates do not
describe the errors associated with the restriction to a
three-dimensional space of SED shapes. An idea of the level of errors
can be gleaned from the comparison of our method to other methods of
$K$-correction. Clearly, $r$, $i$ and $z$ can be reconstructed to the
level of the photometry. Judging from Figure
\ref{k_speck_plot.fitfib.0.1}, the $g$ band can be reconstructed to
about 5\% accuracy. The $u$ band is more difficult to evaluate.
Judging from the poor reconstructions of the $\band{0.3}{g}$ band
(nearly equal to the $\band{0.0}{u}$ band, the reconstructions may be
as bad as 20\% accuracy.

\begin{figure}
\figurenum{\fignum}
\plotone{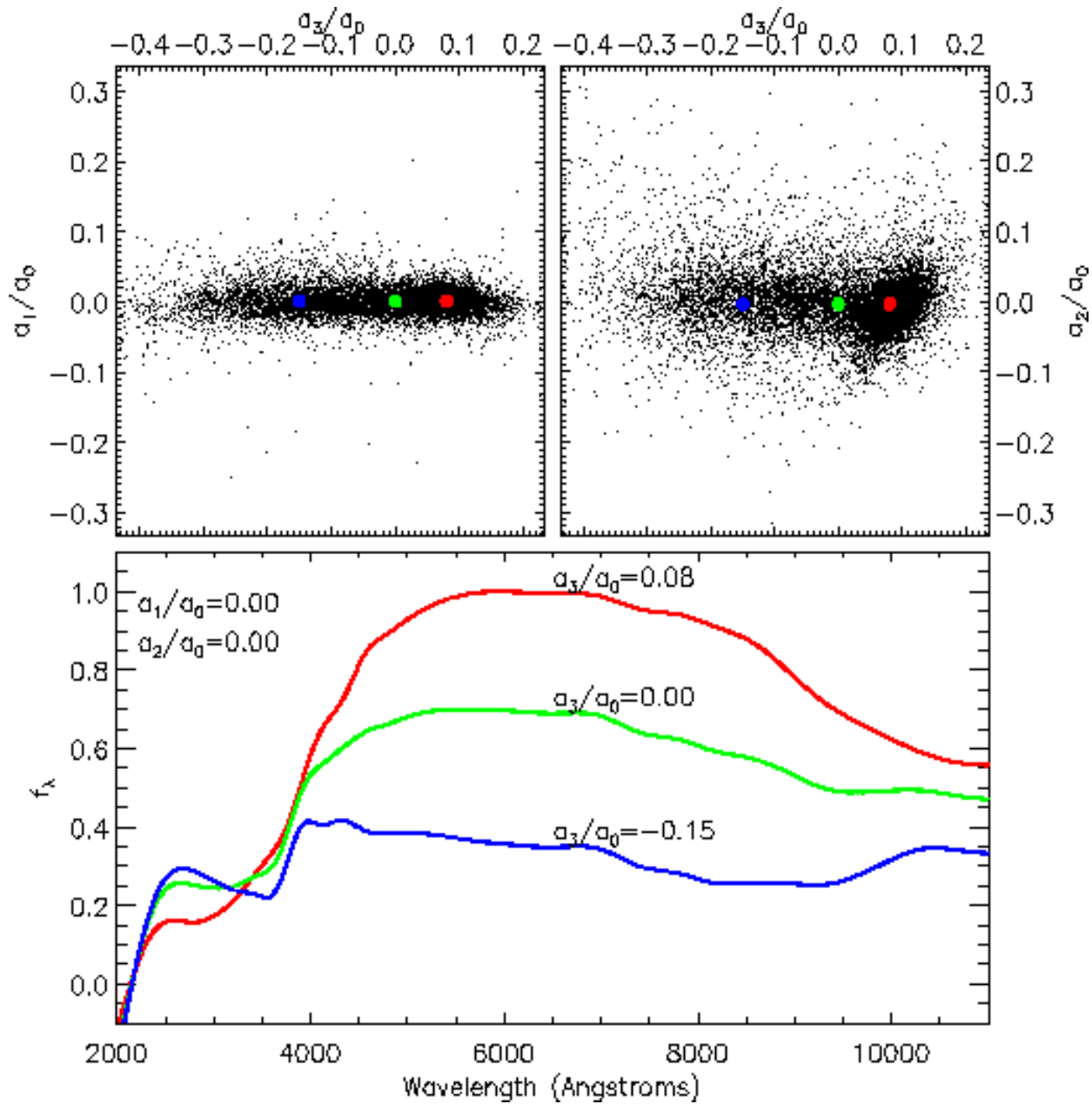}
\caption{\label{k_coeffdist_plot} {\it Top panels}: Distribution of
the components of the four-parameter fit to the five-band SDSS
photometry for a random subsample consisting of 10,000 of the SDSS
galaxies. $a_0$ is linearly proportional to the flux between 3500 \AA
and 7500 \AA, while $a_1$, $a_2$, and $a_3$ contribute no flux in
this range. Thus, the ratios $a_1/a_0$, $a_2/a_0$, and $a_3/a_0$
describe the spectral type of the galaxy. $a_3/a_0$ is the most
variable parameter and thus is the best separator of galaxy
type. The spur extending from the lower left to the upper right from
the red dot in the $(a_3/a_0)$-$(a_2/a_0)$ plane is due to a
degeneracy for galaxies at $z\sim 0.3$, described in detail in Section
\ref{grgap}. {\it Bottom panel}: At fixed $a_1/a_0$ and $a_2/a_0$, the
inferred spectra corresponding to various values of $a_3/a_0$. Near
$a_3/a_0=-0.20$, the spectrum is similar to that of an elliptical
galaxy. For higher values, the spectrum becomes bluer. }
\end{figure}

Topics we have not discussed here are dust- and evolution-correction
of the magnitudes. Some of the scatter in the three-dimensional space
describing the shape of the galaxy SED is probably due to dust. It may
be possible to evaluate the effects of reddening in this space and, by
assuming that galaxies corrected for internal reddening live in an
even lower dimensional space, perform a reddening correction. We will
be investigating this question in the near future. One can perform a
similar exercise with evolution estimates, relating the position of a
galaxy along the $a_1$ axis in Figure \ref{k_coeffdist_plot} to a
particular star-formation history, and estimating the evolution of the
object from that history.

The distribution of galaxies in the three-dimensional space of galaxy
SEDs described here may be useful for other purposes, as well. For
example, you can calculate photometric redshifts using the method
described here. {\tt kcorrect \kversion} has a simple (and very fast)
photometric redshift estimator with very tight core of residuals
($\sigma \sim 0.05$) in a comparison to SDSS galaxies with redshifts.
However, we warn the reader that before this code can be used as a
reliable photometric redshift indicator, more work has to be done
(which the reader is invited to do using the released code) to
identify galaxies which are likely to be redshift outliers and to
handle galaxies at $z>0.5$ (which have photometric observations which
are bluer in the rest frame than for any galaxies in our training
set). Until this method is perfect and our templates are better
constrained, we recommend the photometric redshift method and results
of Csabai {\it et al.}, in preparation. Another use of the
three-dimensional space of galaxy SEDs is to create mock samples of
galaxies observed in any bandpass at any redshift. We are working on
quantifying the correlations between these coefficients and
luminosity, surface-brighntess and radial profile in order to create
mock catalogs which truly reflect the correlations between galaxy
properties found in the data.

\acknowledgments

Special thanks to Ivan Baldry for pointing out an error in this
manuscript and in the distributed code for versions {\tt v1\_10} and
previous.  MB and DWH acknowledge NASA NAG5-11669 for partial support.
MB was supported at the beginning of this work by the DOE and NASA
grant NAG 5-7092 at Fermilab. He is also grateful for the hospitality
of the Department of Physics and Astronomy at the State University of
New York at Stony Brook, who kindly provided computing facilities on
his frequent visits there.

Funding for the creation and distribution of the SDSS Archive has been
provided by the Alfred P. Sloan Foundation, the Participating
Institutions, the National Aeronautics and Space Administration, the
National Science Foundation, the U.S. Department of Energy, the
Japanese Monbukagakusho, and the Max Planck Society. The SDSS Web site
is {\tt http://www.sdss.org/}.

The SDSS is managed by the Astrophysical Research Consortium (ARC) for
the Participating Institutions. The Participating Institutions are The
University of Chicago, Fermilab, the Institute for Advanced Study, the
Japan Participation Group, The Johns Hopkins University, Los Alamos
National Laboratory, the Max-Planck-Institute for Astronomy (MPIA),
the Max-Planck-Institute for Astrophysics (MPA), New Mexico State
University, Princeton University, the United States Naval Observatory,
and the University of Washington.

\newpage

\clearpage
\clearpage

\setcounter{thefigs}{0}

\clearpage
\stepcounter{thefigs}

\clearpage
\stepcounter{thefigs}


\clearpage
\stepcounter{thefigs}

\clearpage
\stepcounter{thefigs}

\clearpage
\stepcounter{thefigs}

\clearpage
\stepcounter{thefigs}
\begin{figure}
\figurenum{\fignum}
\plotone{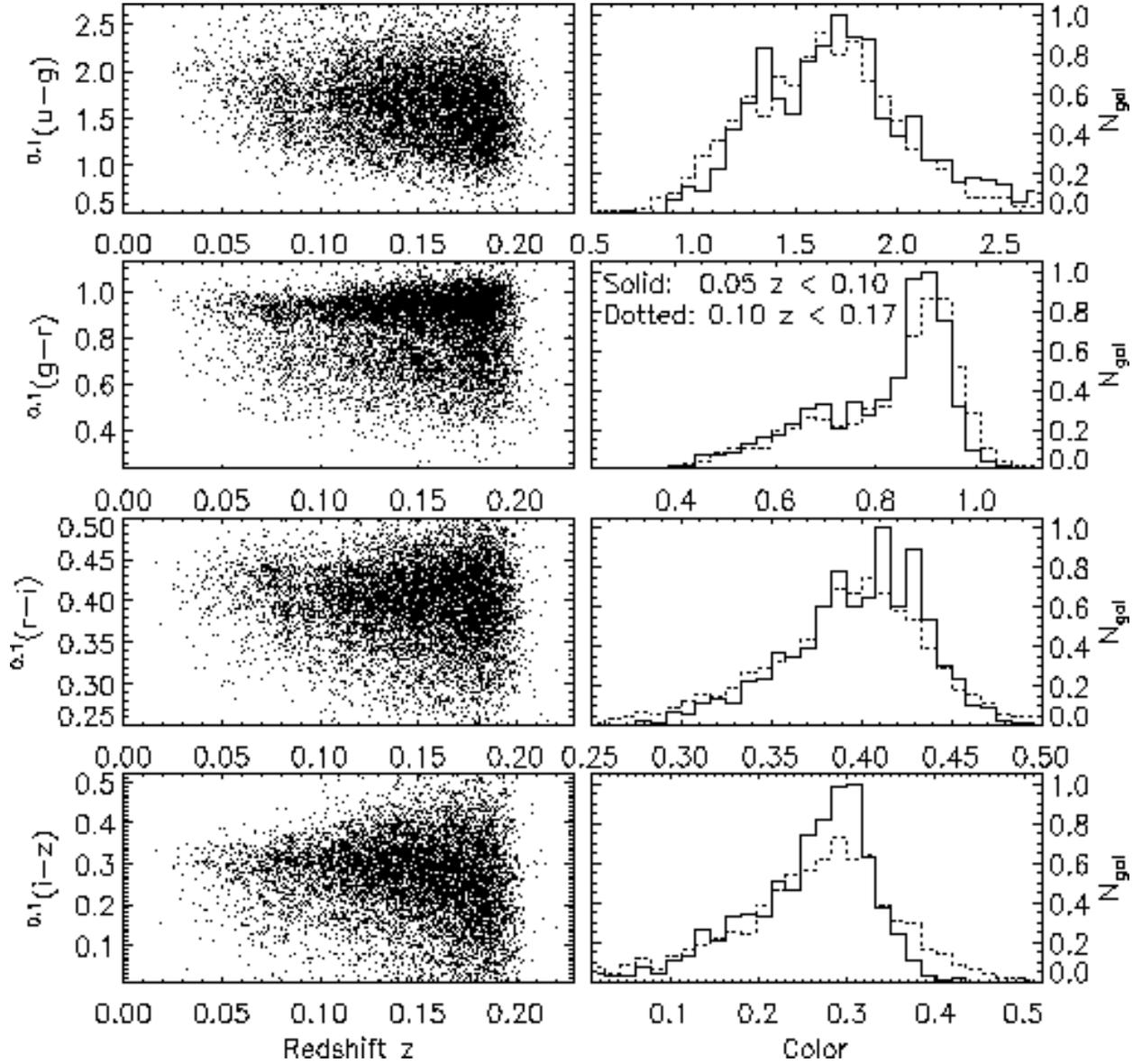}
\caption{\label{main_colors_plot} Similar to Figure
\ref{main_colors_plot.z}, but now the colors are $K$-corrected to
$z=0.1$.  There is very little dependence of the colors on redshift,
even for the $\band{0.1}{u}$ band, where the low-redshift end is an
extrapolation of the data.}
\end{figure}

\clearpage
\stepcounter{thefigs}

\clearpage
\stepcounter{thefigs}

\clearpage
\stepcounter{thefigs}

\clearpage
\stepcounter{thefigs}

\clearpage
\stepcounter{thefigs}

\clearpage
\stepcounter{thefigs}

\clearpage
\stepcounter{thefigs}

\clearpage
\stepcounter{thefigs}


\begin{thebibliography}{DUM}
\bibitem[Binney \& Merrifield (1998)]{binney98a}
Binney, J., \& Merrifield, M.~1998, Galactic Astronomy (Princeton:
Princeton University Press)
\bibitem[Bruzual \& Charlot (1993)]{bruzual93a}
Bruzual, A.~G.,~\& Charlot, S.~1993, \apj, {405}, 538
\bibitem[Bud\'avari {\it et al.} (2000)]{budavari00a}
Budav\'ari, T.; Szalay, A. S.; Connolly, A. J.; Csabai, I.; Dickinson,
M.~(2000), \aj, 120, 1588
\bibitem[Csabai {\it et al.}~(2000)]{csabai00a}
Csabai, I., Connolly, A.~J., Szalay, A.~S., \& Budav\'ari,
T.~2000, \aj, 119, 69
\bibitem[Eisenstein {\it et al.}~(2001)]{eisenstein01a}
Eisenstein, D.~J., {\it et al.}~SDSS Collaboration~2001, 122, 2267
\bibitem[Fan (1999)]{fan99a}
Fan, X.~1999, \aj, 117, 2528
\bibitem[Frei \& Gunn (1994)]{frei94a}
Frei, Z., \& Gunn, J.~E.~1994, \aj, 108, 1476
\bibitem[Fukugita {\it et al.}~(1996)]{fukugita96a}
Fukugita, M., Ichikawa, T., Gunn, J.~E., Doi, M., Shimasaku, K., \&
Schneider, D.~P.~1996, \aj, 111, 1748
\bibitem[Fukugita, Shimasaku, \& Ichikawa (1995)]{fukugita95a}
Fukugita, M., Shimasaku, K., \& Ichikawa, T.~1995, \pasp, 107, 945
\bibitem[Gunn {\it et al.}~(1998)]{gunn98a}
Gunn, J.~E., Carr, M.~A., Rockosi, C.~M., Sekiguchi, M., {\it et al.}~1998, \aj, 116, 3040
\bibitem[Hogg (1999)]{hogg99a}
Hogg, D.~W.~1999, astro-ph/9905116 
\bibitem[Oke \& Gunn (1983)]{oke83a}
Oke, J.~B., \& Gunn, J.~E.~1983, \apj, 266, 713
\bibitem[Oke \& Sandage (1968)]{oke68a}
Oke, J.~B., \& Sandage, A.~1968, \apj, 154, 21
\bibitem[Petrosian (1976)]{petrosian76a}
Petrosian, V.~1976, \apj, 209, L1
\bibitem[Press {\it et al.}~(1992)]{press92a}
Press, W.~H., Teukolsky, S.~A., Vetterling, W.~T., \& Flannery,
B.~P.~1992, Numerical Recipes (Cambridge: Cambridge Univ.~Press)
\bibitem[Schlegel, Finkbeiner \& Davis (1998)]{schlegel98a}
Schlegel, D.~J., Finkbeiner, D.~P., \& Davis, M.~1998, \apj, 500, 525
\bibitem[Schlegel {\it et al.} (2002)]{schlegel02a}
Schlegel, D.~J., {\it et al.}~2002, in preparation
\bibitem[Smith {\it et al.}~(2002)]{smith02a}
Smith, J.~A., {\it et al.}~SDSS Collaboration~2002, \aj, 123, 2121
\bibitem[Stoughton {\it et al.} (2002)]{stoughton02a}
Stoughton, C., {\it et al.}~2002, \aj, 123, 485
\bibitem[Strauss {\it et al.} (2002)]{strauss02a}
Strauss, M.~A., {\it et al.}~2002, submitted to \aj
\bibitem[SubbaRao {\it et al.} (2002)]{subbarao02a}
SubbaRao, M., {\it et al.}~2002, in preparation
\bibitem[York {\it et al.}~(2000)]{york00a}
York, D., {\it et al.}~2000, \aj, 120, 1579
\end{thebibliography}
\end{document}